\documentstyle[12pt]{article}
\begin{document}
\newcommand{\be}{\begin{equation}}
\newcommand{\ee}{\end{equation}}
\newcommand{\ba}{\begin{eqnarray}}
\newcommand{\ea}{\end{eqnarray}}
\begin{center}
{\large \bf Discrete scale invariance and complex dimensions}

\vskip 1cm
Didier Sornette$^{1,2}$ \\

$^{1}$ Laboratoire de Physique de la Mati\`{e}re Condens\'{e}e\\
CNRS and Universit\'{e} de Nice-Sophia Antipolis, Parc Valrose,
06108 Nice, France
\\
$^{2}$  Department of Earth and Space Sciences
and\\ Institute of Geophysics and Planetary Physics\\ University of
California, Los Angeles, California 90095-1567
\end{center}
\vskip 1cm
\vskip 0.5cm
Updated version (Oct. 27, 1998) of the review paper with the same title appeared in
Physics Reports 297, 239-270 (1998)
\vskip 2cm

{\bf Abstract}: We discuss the concept of discrete scale invariance and how it leads to
complex critical exponents (or dimensions), i.e. to the log-periodic corrections to
scaling. After their initial suggestion as formal solutions of renormalization group
equations in the seventies, complex exponents have been studied in the eighties in
relation to various problems of physics embedded in hierarchical systems. Only recently
has it been realized that discrete scale invariance and its associated complex
exponents may appear ``spontaneously'' in euclidean systems, i.e. without the need for
a pre-existing hierarchy. Examples are diffusion-limited-aggregation clusters, rupture
in heterogeneous systems, earthquakes, animals (a generalization of percolation) among
many other systems. We review the known mechanisms for the spontaneous generation of
discrete scale invariance and provide an extensive list of situations where complex
exponents have been found. This is done in order to provide a basis for a better
fundamental understanding of discrete scale invariance. The main
motivation to study discrete scale invariance and its signatures is that it provides
new insights in the underlying mechanisms of scale invariance. It may also be very
interesting for prediction purposes.

\pagebreak

\tableofcontents

\pagebreak
\section{INTRODUCTION}

During the third century BC, Euclid and his students introduced the concept of space dimension,
 which can take positive integer values equal to the number of independent directions. 
We have to wait until the second half of the nineteen
century and the twentieth century to witness the generalization of dimensions to
fractional values. The word ``fractal'' is coined by Mandelbrot \cite{Mandel} to
describe sets consisting of parts similar to the whole, and which can be described by a
fractional dimension (see \cite{Edgar} for a compilation of the most important reprints
of mathemacatical works leading to fractals). This generalization of the notion of a dimension
from integers to real numbers reflects the conceptual jump from
translational invariance to continous scale invariance. 

The goal of this paper is to review the mathematical and physical meaning 
of a further generalization, wherein the dimensions or exponents are taken 
from the set of complex numbers {\footnote{A further generalization to the set of
quaternions (the unique non-commutative generalization of complex numbers on the set
of real numbers) does not bring any new structure.}} (see Ref. \cite{revue} for a
recent review).
We will see that this generalization captures
the interesting and rich
phenomenology of systems exhibiting discrete scale invariance, a weaker form of scale
invariance symmetry, associated with log-periodic corrections to scaling.

Before explaining what is discrete scale invariance, describing its signatures and
importance and studying its mechanisms, let us present a brief historial perspective.
To our knowledge, the first physical model, where sine log-periodic functions arose,
was a model of shock waves in layered systems constructed by Zababakhin in 1965 
\cite{Zababakhin}. He studied the propagation of shock waves in a continuous medium
with periodically distributed properties and, by following the evolution
of the initial impulse by numerical calculation, constructed a solution that also 
periodically reproduces itself on a variable scale, i.e. a log-periodic self-similar
solution. A few years later, Barenblatt and Zeldovich \cite{BarZEL} considered 
Zababakhin's model as a very natural development of self-similarity, in the
following sense. They point out that the determination of exponents in self-similar
solutions of the second type (i.e. not determined solely by dimension analysis) is deeply
connected to the classical eigenvalue problem for linear differential operators.
The point is that a propagating wave $u(x-ct)$ becomes a self-similar homogeneous
function $U(X/T^c)$ by posing $x = \ln X$, $t= \ln T$, which writes again the fact that
performing a scale transformation is the same as a translation in the logarithm.
The nonlinear shock problem is thus connected to a eigenvalue problem with Block-like wave
solutions, which by a mechanism of exponentiation leads to the observed log-periodicity.
Here, the existence of a preexisting linear periodicity is essential and the creation
of log-periodicity can be seen as an exponentiation operation of the discrete
translational invariance into a discrete scale invariance, similarly to other
mechanisms discussed below for discretized systems \cite{BS}.

Novikov has also pointed out in 1966 that structure factors in turbulence
should contain log-periodic oscillations \cite{Novikov}. Loosely speaking,
if an unstable eddy in turbulent flow typically breaks up into two
or three smaller eddies, but not into $10$ or $20$ eddies, then one can suspect
the existence of a preferable scale factor, hence the log-periodic
oscillations. The interest in log-periodic oscillations has been somewhat revived after the
introduction of the renormalization group theory of critical phenomena. Indeed,
the mathematical existence of such corrections has been discussed quite early
in renormalization group solutions for the statistical mechanics of
critical phase \cite{Jona,Nauen,Nieme}. However, these log-periodic oscillations, which
amount to consider complex critical exponents, were rejected for translationally
invariant systems, on the (not totally correct \cite{SS}) basis that
a period (even in a logarithmic scale) implies the existence of one or several
characteristic scales, which is forbidden in these ergodic systems in the critical
regime. Complex exponents were therefore restricted to systems
with discrete renormalization groups. In the eighties, the search for exact solution
of the renormalization group led to the exploration of models put on hierarchical
lattices, for which one can often obtain an exact renormalization group recursion
relation. Then, by construction as we will show below, discrete scale invariance 
and complex exponents and their log-periodic signature appear. 

Only recently has it been realized that discrete scale invariance and its
associated complex exponents can appear spontaneously, without the need for a
pre-existing hierarchical structure. It is this aspect of the domain that is
the most fascinating and on which we will spend most of our time.

\section{WHAT IS DISCRETE SCALE INVARIANCE (DSI)?}

Let us first recall what is the concept of (continuous) scale invariance\,: in
a nutshell, it means reproducing itself on different time or space scales. More
precisely, an observable ${\cal O}$ which depends on a ``control'' parameter $x$ is
scale invariant under the arbitrary change $x \to \lambda x$ {\footnote{Here, we
implicitely assume that a change of scale leads to a change of control parameter as in
the renormalization group formalism. More directly, $x$ can itself be a scale.}}, if
there is a number $\mu (\lambda)$ such that  
\be 
{\cal O} (x) = \mu {\cal O} (\lambda x) ~~~~.
\label{one}
\ee
Eq.(\ref{one}) defines a homogeneous function and is encountered in the theory of
critical phenomena, in turbulence, etc.
Its solution is simply a power law ${\cal O}(x) = C x^{\alpha}$,
with $\alpha = - {\log \mu \over \log \lambda}$, which can be verified directly by 
insertion.  Power laws are the hallmark of scale invariance as the ratio ${{\cal O}
(\lambda x) \over {\cal O} (x)} = \lambda^{\alpha}$ does not depend on $x$, i.e. the
relative value of the observable at two different scales only depend on the {\em
ratio} of the two scales {\footnote{This is only true for a function of a single parameter.
Homogeneous functions of several variables take a more complex form than (\ref{one}).}}.
This is the fundamental property that associates power laws
to scale invariance, self-similarity {\footnote{Self-similarity is the same notion as
scale invariance but is expressed in the geometrical domain, with application to
fractals.}} and criticality {\footnote{Criticality refers to the state of a system
which has scale invariant properties. The critical state is usually reached by tuning a
control parameter as in liquid-gas and paramagnetic-ferromagnetic phase transitions.
Many driven extended out-of-equilibrium systems seem also to exhibit a kind of
dynamical criticality, that has been coined ``self-organized criticality''
\cite{Bak}.}}. 

Discrete scale invariance (DSI) is a weaker kind of scale invariance according to
which the system or the observable obeys scale invariance as defined above only for
specific choices of $\lambda$ (and therefore $\mu$), which form in general an infinite
but countable set of values $\lambda_1, \lambda_2, ...$ that can be written as
$\lambda_n = \lambda^n$. $\lambda$ is the fundamental scaling ratio. This property can
be qualitatively seen to encode a {\it lacunarity} of the fractal structure
\cite{Mandel}.

Note that, since $x \to \lambda x$ and  ${\cal O}(x) \to \mu {\cal O}(\lambda x)$ is
equivalent to  $y = \log x \to y + \log \lambda$ and  $\log {\cal O}(y) \to \log {\cal
O}(y + \log \lambda)+ \log \mu$, a scale  transformation is simply a translation of
$\log x$ leading to a translation of ${\cal O}$. Continuous scale invariance is thus
the same as continuous translational invariance expressed on the logarithms of the 
variables. DSI is then seen as the restriction of the continuous translational 
invariance to a {\it discrete} translational invariance\,: $\log {\cal O}$ is  simply
translated when translating $y$ by a multiple of a fundamental ``unit''  size $\log
\lambda$. Going from continuous scale invariance to DSI can thus be compared with (in
logarithmic scales) going from the fluid state to the solid state in condensed
matter physics! In other words, the symmetry group is no more the full set of
translations but only those which are multiple of a fundamental discrete generator.

\section{WHAT ARE THE SIGNATURES OF DSI?}

\subsection{Log-periodicity and complex exponents}

We have seen that the hallmark of scale invariance is the existence of power laws. The
signature of DSI is the presence of power laws with {\it complex}
exponents $\alpha$ which manifests itself in data by
log-periodic corrections to scaling. To see this, consider the triadic Cantor set
shown in figure 1. This
fractal is built by a recursive process as follows. The first step consists in dividing
the unit interval into three equal intervals of length ${1 \over 3}$ and in deleting the
central one. In the second step, the two remaining intervals of length ${1 \over 3}$ are
themselves divided into three equal intervals of length ${1 \over 9}$ and their central
intervals are deleted, thus keeping $4$ intervals of length ${1 \over 9}$. The process
is then iterated ad infinitum. It is usually stated that this triadic Cantor set has the
fractal (capacity) dimension $D_0 = {\log 2 \over \log 3}$, as the number of intervals grows as
$2^n$ while their length shrinks as $3^{-n}$ at the $n$-th iteration. 

It is obvious to see that, by construction, this triadic Cantor set is geometrically
identical to itself {\it only} under magnification or coarse-graining by factors
$\lambda_p = 3^p$ which are arbitrary powers of $3$. If you take another magnification
factor, say $1.5$, you will not be able to superimpose the magnified part on the
initial Cantor set. We must thus conclude that the triadic Cantor set does not possess
the property of continuous scale invariance but only that of DSI under the fundamental
scaling ratio $3$. 

This can be quantified as follows.  
Call $N_x(n)$ the number of intervals found at the $n$-th iteration of the
construction. Call $x$ the magnification factor. The original unit interval 
corresponds to magnification 1 by definition. Obviously, when the magnification
increases by a factor 3, the number $N_x(n)$ increases by a factor 2
independent of the particular index of the iteration. The fractal
dimension is defined as
\be
D=\lim_{x\rightarrow\infty}{\log N_x(n)\over \ln x}=
\lim_{x\rightarrow 0}{\ln N_x(n)\over \ln x}
={\log 2 \over \log 3} \approx 0.63~~~~ .
\ee
However, the calculation of a fractal dimension usually makes use of
arbitrary values of the magnification and not only those equal to $x=3^p$ only. 
If we increase
the  magnification continuously from say $x=3^p$ to $x=3^{p+1}$, the numbers of
intervals in all classes jump by a factor of 2 at $x=3^p$, but then remains
unchanged until $x=3^{p+1}$, at which point they jump again by an additional
factor of 2. For $3^p<x<3^{p+1}$, $N_x(n)$ does not change while $x$ increases,
so the measured fractal dimension
$D(x)= {\ln N_x(n)\over \ln x}$ decreases. The value
$D=0.63$ is obtained only when $x$ is a positive or negative power of three.
For continuous values of $x$ one has
\be
N_x(n)=N_1(n) x^D P\left({\log x\over \log 3}\right),
\label{two}
\ee
where $P$ is a function of period unity. Now, since $P$ is a periodic function,
we can expand it as a Fourier series
\be
P\left({\log x\over\log 3}\right)= \sum_{n=-\infty}^\infty  c_n
\exp\left(2n\pi i{\ln x\over\ln 3}\right)~~~~.
\label{three}
\ee
Plugging this expansion back into (\ref{two}), it appears that $D$ is replaced by
an infinity of complex values
\be
D_n= D + n i {2\pi\over\log 3}~~~~.
\label{five}
\ee
We now see that a proper characterization of the fractal
is given by this set of  {\it complex dimensions}
which quantifies not only the asymptotic behaviour of the number of fragments at
a given magnification, but also its modulations at
intermediate magnifications. The imaginary part of the complex dimension is directly
controlled by the prefered ratio $3$ under which the triadic Cantor set is exactly
self-similar. Let us emphasize that DSI refers to discreteness in terms of scales,
rather than discreteness in space (eg like discreteness of a cubic lattice
approximation  to a continuous medium).

If we keep only the first term in the Fourier series in (\ref{three}) and insert in
(\ref{two}), we get 
\be
N_x(n)=N_1(n) x^D \left(1 + 2{c_1 \over c_0} \cos (2n\pi
 {\ln x\over\ln 3})\right)~~~~,
\label{four}
\ee
where we have used $c_{-1} = c_1$ to ensure that $N_x(n)$ is real. Expression
(\ref{four}) shows that the imaginary part of the fractal dimension translates itself
into a log-periodic modulation decorating the leading power law behavior. Notice that
the period of the log-periodic modulation is simply given by the logarithm of the
prefered scaling ratio. This is a fundamental result that we will retrieve in
the various examples discussed below. The higher
harmonics are related to the higher order dimensions. 

It is in fact possible to
obtain directly all these results from (\ref{one}). Indeed, let us look for a solution
of the form ${\cal O} (x) = C x^{\alpha}$. Reporting in (\ref{one}), we get the
equation $1 = \mu \lambda^{\alpha}$. But $1$ is nothing but $e^{i2\pi n}$, where $n$ is
an arbitrary integer. We get then 
\be
\alpha = -{\log \mu \over \log \lambda} + i {2\pi n \over \log \lambda}  ~~~~,
\label{rrrt}
\ee
which has exactly the same structure as (\ref{five}). The special case $n=0$ gives the
usual real power law solution corresponding to fully continuous scale invariance. In
contrast, the {\it more general} complex solution corresponds to a possible DSI with
the prefered scaling factor $\lambda$. The reason why (\ref{one}) has solutions in terms 
of complex exponents stems from the fact that a finite rescaling has been done by the
finite factor $\lambda$. In critical phenomena presenting continuous scale invariance, 
(\ref{one}) corresponds to the linearization, close to the fixed point,
 of a renormalization group equation describing
the behavior of the observable under a rescaling by an arbitrary factor $\lambda$.
The power law solution and its exponent $\alpha$ must then not depend on the
specific choice of $\lambda$, especially if the rescaling is taken infinitesimal, 
i.e. $\lambda \to 1^+$.
In the usual notation, if $\lambda$ is noted $\lambda = e^{a_x \ell}$, this implies that 
$\mu = e^{a_{\phi} l}$ and $\alpha = - {a_{\phi} \over a_x}$ is independent of the rescaing 
factor $\ell$. In this case, the imaginary part in (\ref{rrrt}) drops out.

\subsection{Higher order log-periodic harmonics}

Complex critical exponents can be derived from
a discrete renormalization group. From it, we can obtain the relative amplitudes of
the harmonics of the Fourier series expansion of the log-periodic
function, which gives the leading correction to scaling. We can also
discuss the effect of a first non-linear term in the flow map in the
generation of log-periodic accumulation of singular points.

A  naive way of obtaining discrete scale covariance
is to consider a discrete fractal for which only discrete renormalizations are allowed.
Calling $K$ the coupling (e.g. $K=e^{J/T}$ for a spin model)  and $R$
the renormalization group map between two successive generations of the
discrete fractal, one has
\be
\label{rgdis}
f(K)=g(K)+{1\over\mu}f[R(K)]~,
\ee
where $f$ is the free energy per lattice site or bond, $g$ is a regular part
which is made of the free energy of the degrees of freedom summed over between
two successive renormalizations, ${1\over\mu}$ is the ratio of the number
of degrees of freedom between two successive renormalizations. The equation
(\ref{rgdis} is solved recursively by
\be
\label{sol}
f(K)=\sum_{n=0}^\infty {1\over\mu^n}g[R^{(n)}(K)]~,
\ee
where $R^{(n)}$ is the $n^{th}$ iterate of the renormalization transformation.
For general ferromagnetic systems, $R$ has two stable fixed points
 corresponding to $K=0$ and $K=\infty$ fixed points, and an unstable
fixed point at $K=K_c$ which
corresponds to the usual critical point. It is easy to see that $f$ is
singular at $K_c$. Call $\lambda$ the slope of the RG transformation
at $K_c$ ($\lambda>1$ since the fixed point is unstable). Then the $i^{th}$
term in the series for the $k^{th}$ derivative
of $f$ at the fixed point will be proportional to $(\lambda^k/\mu)^i$, giving
rise to a divergence of the series for the $k^{th}$ derivative of $f$ for $k$
large enough, hence the singular behaviour. Now assume that $f\propto |K-K_c|^m$ close to
the critical point. Plugging this form in (\ref{rgdis} gives the constrain,
since $g$ is a regular function,
\be
\label{expo} 
{\lambda^m\over\mu}=1~.
\ee
This equation has an infinity of solutions given by
\be
\label{infsol} 
m_l=m+il\Omega~,
\ee
with
\be
\label{quantit}
m={\ln\mu\over\ln\lambda},\ \Omega={2\pi\over\ln\lambda}~,
\ee
and $l$ is an arbitrary integer.
Decorating the main algebraic behaviour,
there is thus the possibility of terms of the
form $|K-K_c|^m \cos(l \Omega \ln|K-K_c|+\phi_l)$, which are the
looked for log-periodic corections. Such terms
are not generally scale covariant. If continuous changes of scale were
allowed, they would  be incompatible with scale covariance, and thus
forbidden. It is only because the changes of scale are discrete
that they survive  here\,:  only when the change of
scale is of the form $|K-K_c| \to \lambda^l|K-K_c|$ do they
transform multiplicatively, ensuring scale covariance.

To find out more about these correction terms, it is very convenient to use
the Mellin transform as done in \cite{DIL,SS}. Introduce therefore for any function
$f$ the
Mellin transform 
\be
\hat{f}(s)\equiv \int_0^\infty x^{s-1}f(x) dx~.
\ee
We then consider the free energy in (\ref{sol}) in the linear approximation,
replacing $R^{(n)}(K)$ by $K_c+\lambda^n (K-K_c)$. Setting $x=K-K_c$
we have then, by
applying this transformation to both sides of (\ref{sol})
\be
\label{mellin}
\hat{f}(s)=\hat{g}(s){\mu\lambda^s\over \mu\lambda^s-1}~.
\ee
We then reconstruct the original function by taking the inverse
transform
\be
f(x)={1\over 2i\pi}\int_{c-i\infty}^{c+i\infty}\hat{f}(s)x^{-s} ds ~.
\ee
The usefulness of the Mellin transform is that the power law behavior
springs out
immediately from the poles of $\hat{f}(s)$, using Cauchy's theorem.
For a general statistical mechanics model, $g$ being the regular part of
the free
energy has generally the form of the logarithm of a polynomial in $x$.
Factorizing the polynomial,
we do not lose generality by considering $g$ given by
\be
g(x)=\ln(1+x)~,
\ee
for which
\be
\hat{g}(s)={\pi\over s\sin s\pi}~.
\ee
In inverting the Mellin transform, we have two types of poles. The poles
of $\hat{g}$ occur for $s=-n$, $n>0$, and contribute only to the regular part
of $f$, as expected since $g$ is a regular
contribution. The poles of the prefactor in (\ref{mellin}), which
stem from the infinite sum over successive embeddings of scales, occur at
\be
\label{otherpol}
s=-m_l ~,
\ee
with $m_l$ as in (\ref{infsol}. They correspond exactly to the singular contributions
discussed above. Their amplitude is obtained by applying Cauchy's theorem
and is of the order of 
\be
\left|{1\over m+li\Omega}{1\over\sin\pi (m+li\Omega )}\right|~,
\ee
which behaves at large $l$ as $e^{-l\Omega\pi}$. Hence the amplitude
of the log-periodic corrections decays exponentially fast as a function of
the order $l$ of the harmonics. This explains the fact
that mostly the first harmonic accounts for experimental data.

Of course the previous computation suffers from the linear approximation, which
becomes
deeply incorrect as $n$ gets large in (\ref{sol}), hence in the region determining the
singularity.
As discussed in \cite{DIL}, the crucial property missed by the linear approximation
is that $f$ is analytic only in a sector $|arg x|<\theta$ while we treated it
as analytic in the cut plane $|arg x|<\pi$. The true asymptotic decay
of the amplitudes of successive log-periodic harmonics is therefore
slower than initially found, and goes as $e^{-l\Omega\theta}$. The angle
$\theta$ depends specifically on the flow map of the discrete renormalization
group \cite{DIL} and is generally of order $1$. Assuming this
formula $e^{-l\Omega\theta}$ of the decay still gives the right order of
magnitude for $l=1$, we see that the amplitude of the first mode can be much larger than
found above in the linear approximation.

The linear approximation misses more interesting features, discussed in \cite{Derrida}.
Beyond the linear
approximation, the renormalization transformation $R$ can be expanded as a
polynomial in $x$. Let us keep the first two terms only, so $R(x)=\lambda x-\nu x^2$ with
$\nu > 0$. Such a non monotonic RG transformation
occurs for instance in {\bf frustrated} models with both ferromagnetic
and antiferromagnetic interactions. An interesting consequence is that
this transformation has another unstable fixed point at
$x_c={\lambda-1\over\nu}$.
The function $f$ is singular at all the pre-images of this
other fixed point, if the absolute value of the Lyapunov exponent of $R$ at
this fixed point is greater than one namely if $1 < \lambda < 1 + {\nu \over 2}$ or
$\lambda < \lambda_-$ or $\lambda > \lambda_+$ where $\lambda_{\pm} =
{2 + {\nu \over 2} + (4\nu +{\nu^2 \over 4})^{1/2} \over 2}$. This is due to
the fact that these preimages
all go to the fixed point after a finite number of renormalizations.
 Moreover these
preimages accumulate at the original unstable fixed point $x=0$ (figure 1),
 and their
accumulation becomes geometrical very close to $x=0$. When crossing such
a preimage, there is a singularity  in $f$, usually manifested as
a kink in the curve. This sequence of kinks  close to $x=0$
can be fitted well with the  first harmonic of log-periodic oscillations
discussed previously. But the discussion suggests
that there is more to be seen in these oscillations - the geometrical
accumulation of critical points towards the ``main'' critical point.

\subsection{A simple worked-out example\,: the Weierstrass-Mandelbrot fractal function}

Mandelbrot proposed the following extension of the Weierstrass function (in order
to cure the disavantage that the Weiestrass function has a larger scale)
\be
W(t) = \sum_{n=-\infty}^{+\infty} {1-e^{i\gamma^n t} \over \gamma^{(2-D) n}}~e^{i\phi_n}~,
~~~~~~{\rm with}~~1 < D < 2, ~~\gamma>1 ~~~~{\rm and}~~\phi_n~~{\rm arbitrary}~.
\label{wqkjdjk}
\ee
$D$ is the Hausdorf-Besicovich fractal dimension of the graph of $W(t)$ (i.e. the graphs
of the real part or imaginary parts of $W$. We follow \cite{BerryLewis} to present some
important properties that illustrate the previous analysis.

Consider the case where the phases $\phi_n = \mu n$. Then, $W(t)$ obeys the renormalization
group equation
\be
W(\gamma t) = e^{-i\mu}~\gamma^{2-D}~W(t)~.
\label{rrttff}
\ee
This implies that the whole function $W$ can be reconstructed from its value in the
range $t_0 \leq t < \gamma t_0$.

Note that the equation (\ref{rrttff}) does not imply that $W(t)$ is a fractal, because
it is satisfied by smooth functions of the form
\be
f_m(t) = t^{2-D}~\exp [-i (\mu+2\pi m) {\ln t \over \ln \gamma}]~,
\ee
where $m$ is an integer. $W(t)$ can thus be constructed from an infinite series of sum
terms. The Poisson formula 
\be
\sum_{n=-\infty}^{+\infty} f(n) = \sum_{k=-\infty}^{+\infty} \int_{-\infty}^{+\infty}  
f(t) ~e^{2i\pi kt}~ dt   ~ ,	
\ee
also us to obtain this new series as
\be
W(t) = \sum_{m=-\infty}^{\infty} \int_{-\infty}^{+\infty} dn ~
{1-e^{i\gamma^n t} \over \gamma^{(2-D) n}}~e^{i (\mu+2\pi m) n}~.
\ee
After integrating by part, the terms in this series can be expressed as $\Gamma$
functions\,:
\be
W(t) = e^{i{\pi \over 2}  D} ~e^{-{\pi \over 2} \mu \ln \gamma}~
\sum_{m=-\infty}^{\infty} f_m(t)~e^{-\pi m^2/\ln \gamma}~
\Gamma(D-2+i {\mu+2\pi m \over \ln \gamma})~.
\label{tdtdgxgx}
\ee
Application of Stirling's formula shows that the terms decay as $\exp (-2m\pi^2/\ln \gamma)$
as $m\to \infty$ and as $|m|^{-({5 \over 2} - D)}~\exp (icm)$ (where $c$ is a constant) 
as $m \to -\infty$, thus ensuring the convergence of the series. The convergence is fastest
when $\gamma \to 1$ in constrast to (\ref{wqkjdjk}) whose convergence is fastest when 
$\gamma \to \infty$.

Consider the case $\mu = \pi$ leading to
\be
A(t) \equiv {\rm Imaginary~part}~W(t)|_{\mu = \pi} = \sum_{n=-\infty}^{+\infty} ~
{(-1)^n \over \gamma^{(2-D)n}}~\sin(\gamma^n t)~.
\ee
In this case, the term $m=-1$ in (\ref{tdtdgxgx}) 
dominates and gives the following large scale behavior
\be
A(t) \approx - t^{2-D} ~{e^{\pi^2 \over 2 \ln \gamma} \over \ln \gamma}~
{\rm Imaginary~part}~\biggl( e^{i \pi ({D \over 2} + {\ln t \over \ln \gamma}} ~
\Gamma(D-2-i {\pi \over \ln \gamma}) \biggl)~.
\ee
If $\pi/ \ln \gamma$ is sufficiently large, this can be simplified by using
Stirling's formula, to give the trend
\be
A(t) \approx \sqrt{2 \over \ln \gamma} ~({\ln \gamma \over \pi})^{2-D}~~t^{2-D}~
\sin~\biggl( {\pi \over \ln \gamma}~\ln {t \over t_0} + {\pi \over 4} \biggl) ~, 
~~~~~{\rm with}~t_0 = {\pi \over e \ln \gamma}~.
\ee
This is a log-periodic function with a prefered scaling ratio $\gamma$. 
The higher order terms in (\ref{tdtdgxgx}) produces
the non-differentiable fractal structure of the Weierstrass-Mandelbrot function.

This calculation makes explicit the relationship between the main log-periodic
dependence and the higher order fractal structure.

\section{WHAT IS THE IMPORTANCE AND USEFULNESS OF DSI?}

\subsection{Existence of relevant length scales}

Suppose that a given analysis of some data shows log-periodic structures.
What can we get out it them?
First, as we have seen, the period in log-scale of the log-periodicity is directly
related to the existence of a prefered scaling ratio. Thus, log-periodicity must
immediatly be seen and interpreted as the existence of a set of prefered characteristic
scales forming all together a geometrical series $..., \lambda^{-p}, \lambda^{-p+1},
..., \lambda, \lambda^2, ...., \lambda^n, ....$. The existence of such prefered scales
appears in contradiction with the notion that a critical system, exhibiting scale
invariance has an infinite correlation length, hence only the microscopic ultraviolet
cut-off and the large scale infra-red cut-off (for instance the size of the system)
appear as distinguishable length scales. This recovers the fact that DSI is a
property different from continuous scale invariance. In fact, it can be shown
\cite{Wallace} that exponents are real if the  renormalization group is a gradient
flow, a rather common situation for systems at thermal
equilibrium, but as we will see, not the only one by far.
Examples when this is not the case can be found especially in random systems,
out-of-equilibrium situations and irreversible growth problems.
In addition to the existence of a single prefered scaling ratio and its associated 
log-periodicity discussed above, there can be several prefered
ratios corresponding to several log-periodicities that are superimposed. This can lead
to a richer behavior such as log-quasi-periodicity. Quasiperiodicity has been
suggested to describe the scaling properties of diffusion-limited-aggregation clusters \cite{DLA}. 

Log-periodic structures in the data indicate that the system and/or the
underlying physical mechanisms have characteristic length scales. This is extremely
interesting as this provides important constraints on the underlying physics. Indeed,
simple power law behaviors are found everywhere, as seen from the explosion of the
concepts of fractals, criticality and self-organized-criticality \cite{Bak}. For
instance, the power law distribution
of earthquake energies which is known as the Gutenberg-Richter law can be obtained
by many different mechanisms and a variety of models and is thus extremely limited in
constraining the underlying physics. Its usefulness as a modelling constraint is even doubtful, 
in contradiction with the common belief held by physicists on the
importance of this power law. In contrast, the presence of log-periodic features
would teach us that important physical structures, that would be hidden in the fully
scale  invariant description, existed. 
 
 \subsection{Non-unitary field theories}
 
In a more theoretical vein, we must notice that complex exponents do not
appear in the canonical exactly solved models of critical phenomena
like the square lattice Ising model or Bose Einstein condensation. This is
because such models satisfy some sort of unitarity. From conformal
invariance \cite{Cardy}, it is known that the
exponents of two dimensional critical models can be measured as amplitudes of
the correlation lengths in a strip geometry. Since the Ising model transfer
matrix can be written in a form which is symmetric, all its eigenvalues are
real, therefore all its exponents are real. The other standard example
where exponents can be computed is $\epsilon$ expansion. However in that
context,  there is an  attitude, inherited from particle physics, to
think mostly of Minkowski field theories. For instance in axiomatic field
 theory, Euclidian field theories are defined mostly as analytic continuations
of Minkowski field theories. Now
complex exponents, as we have argued \cite{SS}, make perfect sense for
Euclidian field theories, but lead to totally ill-behaved Minkowski
field theories, with exponentially diverging correlation functions.
An approach based on any sort of equivalence between the two points
of view is bound to discard complex exponents (as well say as complex masses).
The complex exponents can thus be viewed as resulting from the breaking of equivalence
(or symmetry under Wick rotation) of the Euclidian and Minkowski field theories.
As it is now undertood that quantum field theories are only effective theories
that are essentially critical {\footnote{The microscopic cut-off is the Planck scale
$\sim 10^{-36} m$ while the macroscopic cut-offs (or correlation lengths) corresponding
to the observed particle masses such as for the electron are of the order of
$10^{-15} m$.   This is a situation where the correlation length is thus $10^{21}$
times larger than the ``lattice'' size, very close indeed to criticality!}}
\cite{Weinberg}, could there be a relation between the spectrum of observed particle
masses and the characteristic scales appearing in DSI and its variants and
generalizations?

\subsection{Prediction}

Lastly, it is important to stress the practical consequence of log-periodic structures.
For prediction purposes, it is much more constrained and thus reliable to fit a part of
an oscillating data than a simple power law which can be quite degenerate especially in
the presence of noise. This remark has been used and is vigorously
investigated in several applied domains, such as earthquakes
\cite{SorSam,SSS1,SSS2,Kobe}, rupture prediction \cite{Anifrani} and financial crashes
\cite{crash1,Freund,crash2}.

\section{SCENARIOS LEADING TO DSI}

After the rather abstract description of DSI given above, we now discuss the physical
mechanisms that may be found at its origin. It turns out that there is not a unique
cause but several mechanisms may lead to DSI. Since DSI is a partial
breaking of a continuous symmetry, this is hardly
surprising as there are many ways to break down a symmetry. 
We describe the mechanisms that have been studied
and are still under investigation. The list of mechanisms is by no mean exhaustive
and other mechanisms may exist. We have however tried to present a rather complete
introduction to the subject. 

It is essential to
notice that all the mechanisms described below involve the existence of a
characteristic scale (an upper and/or lower cut-off) from which the DSI can develop
and cascade. In fact, for characteristic length scales forming a 
geometrical series to be present, it is unavoidable that 
they ``nucleate'' from either a large  size or a small mesh. This remark has the 
following important consequences\,:
even if the mathematical solution of a given problem
contains in principle complex exponents, if there are no such cut-off scales
to which the solution can ``couple'' to, then the log-periodicity will be absent in the 
physical realization of the problem. An example
of this phenomenon is provided by the interface-crack stress singularity described below.

\subsection{Built-in geometrical hierarchy}

The most obvious situation occurs when some physical system is put on a
pre-existing discrete hierarchical system, such as the Bethe lattice, or a fractal tree.
Since the hierarchical system contains by construction a discrete hierarchy of scales
occurring according to a geometrical series, one expects and does find complex exponents
and their associated log-periodic structures. Examples are fractal dimensions of
Cantor sets \cite{Fournier,Fournier2,Orlan}, percolation \cite{Kapitulnik}, ultrametric
structures \cite{Vladmackey}, wave propagation in fractal systems \cite{Bessis},
magnetic and resistive effects on a system of wires connected along the Sierpinski
gasket\cite{Doucot}, Ising and Potts models \cite{DIL,Meurice}, fiber bundle
rupture \cite{Newman,SSS2}, sandpiles \cite{Stanley}. Quasi-periodic and aperiodic 
structures can also often be captured by a discrete renormalization group and can be expected 
to lead to discrete scale invariance and log-periodicity. For instance, the quantum
$XY$ spin chain with quasi-periodic two-valued exchange couplings \cite{LuckNiew}
 has its zero-field
specific heat and susceptibility behaving as power laws of the temperature, modulated
by log-periodic modulations. The zero-temperature magnetization is a Cantor function
of the applied field and the density of state is also a power law of the distance
to gaps with log-periodic modulations. Similarly, log-periodic
corrections to scaling of the amplitude of the surface magnetization 
have been found for aperiodic modulations of the coupling in Ising quantum chains \cite{Karevski}.
Analytical and numerical
calculations associated with succesive hierarchical approximations to
multiscale fractal energy spectra show that, in a certain range of
temperatures, the specific heat displays log-periodic oscillations as a
function of the temperature \cite{Vallejos}. The scattering cross-section of a radiation
scattered off a DSI fractal structure also exhibits log-periodic oscillations as
a function of the scattering wavenumber parameter \cite{Olemskoi}.

Multifractal models of geophysical fields have used 
discrete scale ratios to construct isotropic and anisotropic cloud and rain structures
\cite{Pecknold,Veneziano}.

\subsubsection{Potts model on the diamond lattice}

Let us now give some details to see more clearly how physics on hierarchical systems 
leads to log-periodicity. As a canonical example, we treat the Potts model \cite{Wu} on
the diamond lattice \cite{DIL}. This lattice is  obtained by starting with a bond at
 magnification 1, replacing this bond by four bonds arranged in the
shape of a diamond at magnification 2, and so on, as illustrated figure 2.
At a given magnification $2^p$, one sees $4^p$ bonds, and
thus ${2\over 3}(2+4^p)$ sites.

The spins $\sigma_i$ are located at the vertices of the
diamond fractal. In the same way that
the lattice appears different at different scales from a geometrical point
of view, one sees a different number of spins at different scales, and
they will turn out to interact in a scale dependent way.
For a given magnification $x=2^p$, the spins we can see are
coupled with an interaction energy
\be
E=-J\sum_{<ij>}\delta(\sigma_i\sigma_j),
\label{energy}
\ee
where $J$ is the coupling strength, the sum is taken over nearest neighbors
and the delta function equals one if arguments are equal, zero otherwise
The system is assumed at thermal
equilibrium, and the spin configurations evolve
randomly in time and space in response to thermal fluctuations with a probability
proportional to the Boltzmann factor  $e^{-\beta E}$, 
where $\beta$ is the inverse of the temperature. The partition function $Z$
at a given magnification $x=2^p$ is
$$
Z_p=\sum_{\{\sigma\}}e^{-\beta E}
$$
where the sum is taken over all
possible spin configurations which can be seen at that scale. 
We do not compute $Z_p$ completely, but
first perform a partial summation over the spins seen at one scale
and which are coupled only to two other spins. This is how, in this
particular example, one
can carry out the program of the renormalization group
 by solving a succession of problems at
different scales.
Let us isolate a particular diamond,
call $\sigma_1,\sigma_2$ the spins at the extremities and $s_1,s_2$ the
spins in between as in figure 2. The contribution of this
diamond to $e^{-\beta E}$ is
$$
K^{\delta(\sigma_1,s_1)+\delta(\sigma_2,s_1)+\delta(\sigma_1,s_2)
+\delta(\sigma_2,s_2)},
$$
where we have defined $K=e^{\beta J}$. Since $s_1,s_2$ enter only in this
particular product, we can perform
summation over them first when we compute $Z_p$. The final result depends on
whether
$\sigma_1$ and $\sigma_2$ are equal or different:
\begin{eqnarray}
\sum_{s_1,s_2}K^{\delta(\sigma_1,s_1)+\delta(\sigma_2,s_1)+\delta(\sigma_1,s_2)
+\delta(\sigma_2,s_2)}
&=& (2K+Q-2)^2,\quad\sigma_1\neq\sigma_2 \\
&=& (K^2+Q-1)^2,\quad\sigma_1=\sigma_2~,
\label{inti}
\end{eqnarray}
so we can write
\begin{eqnarray}
\sum_{s_1,s_2}K^{\delta(\sigma_1,s_1) +\delta(\sigma_
2,s_1)+\delta(\sigma_1,s_2) +\delta(\sigma_2,s_2)}\\
\nonumber
&=& (2K+Q-2)^2\left[1+
\left({(K^2+Q-1)^2\over (2K+Q-2)^2}-1\right)\delta(\sigma_1,\sigma_2)\right] \\
\nonumber
&=& (2K+Q-2)^2 K'^{\delta(\sigma_1,\sigma_2)}~,
\label{intii}
\end{eqnarray}
where we used the identity
\be
K'^{\delta(\sigma_1,\sigma_2)}=1+(K'-1)\delta(\sigma_1,\sigma_2)~,
\label{identt}
\ee
and  we set
\be
K'\equiv\left({K^2+Q-1 \over 2K+Q-2}\right)^2~.
\ee
If we perform this partial resummation in each of the four diamonds, we
obtain exactly the system at a lower magnification $x=2^{p-1}$.
We see therefore that the interaction of spins tranforms very simply  when
the lattice is magnified\,: at any  scale,  only nearest neighbor
spins are coupled, with a scale dependent coupling determined recursively
through the {\bf renormalization group map}
\be
K_{p-1}=\left({K_p^2+Q-1 \over 2K_p+Q-2}\right)^2 \equiv\phi(K_p)~.
\label{mappp}
\ee

The spins which are ``integrated out'' by going
from one magnification to the next simply contribute an overall numerical
factor to the partition function,  which is equal to the factor
  $(2K+Q-2)^2$ per edge of (\ref{intii}). Indeed, integrating out the spins $s_1$ and
$s_2$ leaves only $\sigma_1$ and $\sigma_2$ whose interaction weight is by
definition
$K'^{\delta(\sigma_1,\sigma_2)}$, if $K'$ represents the effective
interaction weight
at this lower magnification $2^{p-1}$. The additional numerical factor
shows that the
partition function is not exactly invariant with the rescaling but
transforms according to
\be
Z_p(K)=Z_{p-1}[\phi(K)](2K+Q-2)^{2. 4^{p}}~,
\label{renpart}
\ee
since there are $4^p$ bonds at magnification $2^p$.
Now the free energy, which is defined as the logarithm
of the partition function per bond, reads
$$
f_p(K)={1\over 4^{p+1}}\ln Z_p(K)
$$
From (\ref{renpart}), we deduce the following
\be
f_p(K)=g(K)+{1\over 4}f_{p-1}(K')~,
\label{renfree}
\ee
where
\be
g(K)={1\over 2}\ln(2K+Q-2)~.
\label{ggg}
\ee
For an infinite fractal, the free energy for some
microscopic coupling $K$ satisfies therefore
\be
f(K)=g(K)+{1\over \mu}f(K')~,
\label{freefinal}
\ee
where $\mu=4$. This explicit calculation makes clear the origin of the
scaling for the free energy\,: the interaction weights remain of the same
functional form at each (discrete) level of magnification, up to a
multiplicative factor
which accounts for the degrees of freedom ``left-over'' when integrating
from one magnification to the next. This is the physical origin of the function
$g$ in (\ref{freefinal}).

\subsubsection{Fixed points, stable phases and critical point}

Consider the map $K'=\phi(K)$ (\ref{mappp}). It exhibits three fixed points (defined
by $K'=K=\phi(K)$) located at
$K=1, K=\infty, K=K_c$ where $K_c$ is easily determined numerically, for
instance
$K_c\approx 3.38 $ for $Q=2$, $K_c
\approx 2.62$ for $Q=1$. That $K=1$ and $K=\infty$ are fixed points is  obvious. 
The former corresponds
 to totally uncoupled spins, the latter to spins which are forced to have the
 same value. In both cases, the dynamics disappears completely, and one gets
 back to a purely geometrical problem. Observe that these two fixed points are
 attractive. This means that if we start with some coupling say $K>K_c$ deep
down in the system, that is for very large magnifications, when one diminishes
 the magnification to look at the system at macroscopic scales, spins appear
 almost always parallel, and therefore are more and more correlated as one
reduces magnification.  Similarly if we start with $K<K_c$ spins are less and
less correlated as one reduces magnification. The condition $K>K_c$
together with
the definition $K=e^{\beta J}$ implies $\beta > \beta_c$, i.e. corresponds to
the low-temperature regime dominated by the energy. The physical meaning of
the attraction of
the renormalization group flow to the fixed point $K=\infty$, i.e. zero
temperature, means
that the macroscopic
state of the spins is ferromagnetic with a macroscopic organization where a
majority of
spins have the same value. 
Similarly, the condition $K<K_c$ implies $\beta
< \beta_c$, i.e.
corresponds to  the high-temperature regime dominated by the entropy or
thermal agitation. The
physical meaning of the attraction of the renormalization group flow to the
fixed point $K=0$,
i.e. infinite
temperature, means that the macroscopic state is completely random with
zero macroscopic
magnetization. 

The intermediate fixed point
$K_c$, which in contrast is repulsive, plays a completely different and
very special role.
It does not describe a stable thermodynamic phase but rather the transition
from one phase
to another. The repulsive nature of the renormalization group map flow means
that this
transition occurs for
a very special value of the control parameter (the temperature or the
coupling weight
$K=K_c$). Indeed, if we have spins interacting with a coupling strength right
at
$K_c$ at microscopic scales, then even by reducing
 the magnification we still see spins interacting with a coupling strength
right
at $K_c$! This is also a point
 where spins must have an infinite correlation length (otherwise it would
 decrease to zero as magnification is reduced, corresponding to a different
effective interaction): by definition it is a {\bf critical } point.

 Close to $K_c$ we can
 linearize the renormalization group transformation
\be
K'-K_c\approx \lambda(K-K_c)~,
\label{linearizi}
\ee
where $\lambda = {d\phi\over dK}|_{K_c}>1$. For couplings
close enough to the critical point, we now see that as we increase
 magnification, the change in coupling becomes also very simple\,; only,
it is not the coupling that gets renormalized by a multiplicative factor,
 but the distance to $K_c$. 
 
 The equation (\ref{freefinal}) together with (\ref{linearizi}) 
provides an explicit realization of the postulated functional form (\ref{one}) 
(up to the non-singular term $g$), where the coupling parameter $K$ (in fact
$K-K_c$) plays the role of the
control parameter $x$.

\subsubsection{Singularities and log periodic corrections}

The renormalization group equations
(\ref{mappp}) and (\ref{freefinal}) can be solved for the free energy by
\be
f(K)=\sum_{n=0}^\infty {1\over \mu^n}g[\phi^{(n)}(K)]~,
\label{solsol}
\ee
where $\phi^{(n)}$ is the $n^{th}$ iterate of the transformation $\phi$ (eg
$\phi^{(2)}(x)=\phi[\phi(x)]$). Now it
is easy to show \cite{Derrida} that the sum (\ref{solsol}) is
{\bf singular} at $K=K_c$. This stems from the fact that $K_c$ is an unstable fixed point,
so the derivative of $\phi$ at $K_c$ is $\lambda>1$. Therefore if we consider
the $k^{th}$
derivative of $f$ in (\ref{solsol}) it is determined by a series whose
generic term behaves as $\left({\lambda^k\over\mu}\right)^n$ which is greater
than 1 for $k$ large enough, so this series diverges. In other words
high enough  derivatives of $f$  are  infinite at $K_c$. Very generally,
this implies that close to $K_c$ one has
\be
f(K)\propto (K-K_c)^m~,
\label{singi}
\ee
where $m$ is called a {\bf critical exponent}. For instance if $0<m<1$, the
derivative
of $f$ diverges at the critical point.  Plugging this back in (\ref{freefinal}), we
see that, since $g$
is regular at $K_c$ as can be checked easily from (\ref{ggg}), we can substitute it
in (\ref{freefinal}) and recover the leading critical behavior and derive
$m$ solely from the equation
$(K-K_c)^m={1\over\mu}[\lambda(K-K_c)]^m$ involving only the singular part,
with the flow map which has been linearized in the vicinity of the critical point.
Therefore, the exponent satisfies $\lambda^m=\mu$, an equation
that we have already encountered and whose general
solution is given by 
\be
m_n={\ln\mu\over\ln\lambda}+ni {2\pi\over\ln\lambda}~.
\label{expoqfq}
\ee
To get expression (\ref{expoqfq}), we have again used the identity $e^{i2\pi n} = 1$.
We see that because there is discrete scale invariance (namely (\ref{freefinal})
holds which relates the free energy only at two different scales in the ratio $2$),
 nothing forces $m$ to
actually be a real number. In complete analogy with the case of complex
fractal dimension, a critical phenomenon on a fractal exhibits {\bf complex
critical exponents}.  Of course $f$ is real, so the most general form
of $f$ close to the critical point should be
\be
f(K)\approx(K-K_c)^m\left\{a_0+\sum_{n>0}a_n\cos[2\pi n\Omega\ln(K-K_c)+\Psi_n]\right\}~,
\label{genefff}
\ee
where
\be
m={\ln\mu\over\ln\lambda},\quad \Omega={1\over\ln\lambda}~,
\label{defsss}
\ee
hence exhibiting the log-periodic corrections. Derrida et al. \cite{DIL} have
studied this example more quantitatively and find that the amplitude of the 
log-periodic oscillations are of the order of $10^{-4}$ times less that the 
leading behavior. This is thus a small effect. In contrast, the examples below 
exhibit a much stronger amplitude of the log-periodic corrections to scaling, 
that can reach $10\%$ or more.

\subsubsection{Related examples in programming and number theory}

Log-periodicity, many of which are of a fractal nature, are found in the solutions of
algorithms based on a recursive {\it divide-and-conquer}
strategy \cite{Flajolet} such as heapsort, mergesort, Karatsuba's multiprecision
multiplication, discrete Fourier transform, binomial queues, sorting networks, etc.
For instance, it is well-known that  
the worst time cost measured in the number of comparisons that are
required for sorting $n$ elements by the MergeSort procedure is given by $n\log_2 n$
to leading order. It is less known that the first subleading term is 
$n P(\log_2 n)$, where $P$ is periodic \cite{Flajolet}.

Reducing a problem to number theory is like striping it down to its sheer
fundamentals. In this vein, arithmetic functions related to the number representation
systems exhibit various log-periodicities. For instance, 
the total number of ones in
the binary representations of the first $n$ integers is ${1 \over 2} n \log_2 n + n
F(\log_2 n)$, where F is a fractal function, continuous, periodic and nowhere
differentiable \cite{Delange}.

The statistical distribution of energy level spacings in two-dimensional harmonic
oscillators with irrational frequency ratio $\omega_1/\omega_2$ also exhibit
a discrete hierarchical structure deeply connected to the properties of irrational
numbers. This problem arises in the study of the correspondence between classical
and quantum dynamics in Hamitonian systems \cite{Gutzwiller}. For integrable classical
systems, the generic level spacing distribution is the Poisson's law of exponential decay
with a maximum at zero, corresponding to level clustering. The two-dimensional harmonic
oscillator problem, possibly the simplest integrable system, does not follow this generic
rule. The spectrum is found to exhit a discrete and rigid structure controlled by the
continued fraction expansion of the irrational number $\omega_1/\omega_2$ \cite{Whan}. 
Following the Einstein-Brillouin-Keller (EBK) semi-classical quantization rule, which for
harmonic oscillators give the exact quantum mechanical results, the energy levels are
\be
E_{m_1, m_2} = {h \over 2\pi}~ [(m_1 + {1 \over 2}) \omega_1 + (m_2 + {1 \over 2}) \omega_2]~.
\ee
In units of $h/ 2\pi$, we see that the difference between two energy levels is
$(m_1 -m'_1) \omega_1 + (m_2 -m'_2) \omega_2$. For $\omega_1/\omega_2$ irrational, this
is never zero. Consider first the simplest and best irrational, the golden mean 
$\sigma_1 = \omega_1/\omega_2 = (\sqrt{5} - 1)/2$ which has the continued fraction expansion
$\sigma_1 = [1, 1, 1, ...]$. The closest difference are obtained for 
$m_1 -m'_1$ and $m'_2 -m_2$ being the successive Fibonacci numbers ($F_0=F_1=1, 
F_{n+1} = F_N + F_{n-1}$) which are such that 
$F_n/F_{n+1}$ are the best rational approximations to $\sigma_1$. Due to their property and that
of the golden mean, we have $F_{n-1} - \sigma_1 F_n = (-\sigma_1)^{n+1}$. Thus all level
spacings are integer powers of $\sigma_1$. The entire distribution of level spacings 
$\Delta E$ satisfies
\be
P_{\sigma_1}(\Delta E) \sim \Delta E^{-2}~ \delta(\Delta E - \sigma_1^n)~, ~~~~{\rm for}~n=1, 2, ...
\ee
We recognize a special case of a discrete scale invariant distribution with prefered scaling
ratio $\sigma_1$.

This results holds for other irrational numbers. Consider $\omega_1/\omega_2 = \sigma_2 = 
\sqrt{2} - 1 = [2, 2, 2, ....]$. Now, $P_{\sigma_2}(\Delta E)$ not only peaks at
$\Delta E = |G_n \sigma_2 - G_{n-1}$ corresponding to the continued fraction
approximations $G_{n-1}/G_n$ (with $G_0=G_1=1$ and $G_{n+1} = 2G_n + G_{n-1}$), it also
shows peaks that correspond to the so-called intermediate fractions 
$(G_n + G_{n-1})/(G_{n+1}+G_n)$ \cite{fractionexp}. These intermediate fractions are the 
second best approximations. The peak corresponding to the intermediate fraction 
$(G_n + G_{n-1})/(G_{n+1}+G_n)$ has
the same height as the one corresponding to $G_{n-1}/G_n$, while these main peaks continue to 
follow the inverse-square law. 

For a general ratio $\omega_1/\omega_2 = 
[a_1, a_2, a_3, ...., a_n, ...]$, it is best approximated by the successive fractions 
$p/q = [a_1, a_2, a_3, ...., a_n]$ obtained by successive truncations of the continued
fraction expansion. In addition, when $a_n>1$, there are $a_n-1$ intermediate fractions
$(p_{n-1} + kp_n)/(q_{n-1}+kq_n)$ with $k=1, 2, ... , a_n-1$, between $p_{n-1}/q_{n-1}$
and $p_n/q_n$ which also provide a good approximation to $\omega_1/\omega_2$. There is
a one-to-one correspondence between the allowed nearest neighbor level spacings and the 
continued and intermediate fraction approximations of the frequency ratio $\omega_1/\omega_2$
\cite{onetoone}. In the case where $\omega_1/\omega_2$ belongs to the class of relatively
simple irrationals known as quadratic numbers, for which the continued fraction expansion
is periodic (which includes $\sigma_1$ and $\sigma_2$), the entire distribution 
$P_{\omega_1/\omega_2}(\Delta E)$ is self-similar and log-periodic\,: 
$\Delta E^2~P_{\omega_1/\omega_2}(\Delta E)$ is periodic in a log-log scale with period
given by the logarithm of the prefered scaling ratio $\omega_1/\omega_2$. This log-periodicity
is obviously related to the fact that the continued fraction expansion of a quadratic 
irrational number is periodic. Since going from one level to the next in the continued
fraction expansion is a multiplicative operation, the periodicity in the continued fraction
expansion is reflected in a log-periodicity in the number properties.

The exact log-periodicity is lost for more general irrational numbers while the general scaling
inverse square law seems to persist. In addition, local log-periodicity can be observed 
resulting from local periodicity of the continued fraction expansion of the irrational 
number. In sum, log-periodicity arises in this problem because one looks for an 
extremal property, namely the successive spacings, in the presence of a
periodic continued fraction expansion.

\subsection{Diffusion in anisotropic quenched random lattices}

In this scenario, the DSI hierarchy is constructed dynamically in a random walk
process due to intermittent encounters with slow regions \cite{BS}. Consider
a random walker jumping from site to site. Bonds between sites are of two types\,: (i)
directed ones on which the walker surely goes from his site to the next on his right
(``diode'' situation)\,; (ii) two-way bonds characterized by a rate $u$ (resp. $v$) 
to jump to the neighboring site on his right (resp. left). The fraction of two-way bonds
is $1-p$ and the fraction of directed bonds is $p$. We construct a {\it frozen} random
lattice by choosing a given configuration of randomly distributed mixtures of the
two bond species according to their respective average concentration $p$ and $1-p$.
The exact solution of this problem has been given in \cite{BS} and shows very clearly
nice log-periodic oscillations in the dependence of $\langle x^2 \rangle$ as a function
of time, as seen in figure 3.

We now present a simple scaling argument \cite{SSS2} which recovers the exact results.
To do so, we assume ${u\over v}<<1-p<<1$. We are thus in a situation where
most bonds are directed and dilute clusters of two-way bonds are present. In addition, 
the two bonds are strongly impeding the progress of the walker as the forward rate to
the right is much smaller than the backward rate to the left.

In this situation, the random walker progresses at constant velocity to the
right as long as it encounters only diode bonds and gets partially trapped when it
encounters two-way bonds. To see how DSI is spontaneously generated, we estimate the 
typical number of jumps $\tau_k$ needed for the random walker to pass $k$ adjacent
two-way bonds, i.e. a connected cluster of $k$ two-way bonds. In the limit
${u \over v}<<1$, $\tau_k \sim \left({v \over u}\right)^k$ {\footnote{and not 
$\tau_k \sim k {v\over u}$. This is due to the fact that the walker goes back many
times before escaping from the cluster of size $k$.}}. Using the fact
that the average separation between $k$-tuples of two-way bonds is approximately
$(1-p)^{-k}$, if ${u\over v} << (1-p)$, the typical number of jumps needed for
the random walker to go beyond the
first $k$-tuple of consecutive two-way values is completely
 dominated by $\tau_k$.
One thus expects the rate as a function of the number of jumps
 to exhibit local minima at $\tau =
\tau_k$: these are the jump-numbers time scales. The second part of the scaling
argument consists in recognizing that the random walker has to cover a typical distance
from the origin to encounter the first k-tuple
 of consecutive `two-way' values, of the order of $(1-p)^{-k}$. In the limit
where ${u\over v}<<(1-p)<<1$, we can thus write the approximate
renormalization group equation
\be
x(\tau) \approx (1-p) x(\lambda\tau) +g(\tau)~~~~,
\ee
where we have set $\lambda \equiv {v\over u}$ and $g(\tau)$ is some regular
function taking into account various local effects that correct the main scaling.
Notice the similarity with (\ref{freefinal}).
Because this renormalization group equation can be written only at scales which
are powers of $\lambda$, we are back to the situation discussed before. We
see in particular that $x \sim \tau^{\nu}$ with 
\be
\nu={\log(1-p)\over \log(u/v)} + i {2\pi n \over  \log(u/v)}~.
\label{ertdff}
\ee
This result for the exponent $\nu$ turns out to be exact.
Furthermore, the range of parameters over which this holds is
much larger than suggested by this intuitive argument. More precisely, as soon as
${u\over v} < (1-p)$, one finds $x(\tau) \sim \tau^{\nu}
P({  \log \tau \over \log (v/u)})$, where $\nu = {\log(1-p)\over \log(u/v)}$ and  $P$ is
a periodic function of unit period. This prediction is remarkably well-confirmed by
numerical simulations and recovers the exact calculation of \cite{BS}. This is shown in 
figure 3.

These log-periodic oscillations are not artifacts of 1D-systems as
random walks with a fixed bias direction on randomly diluted 3D-cubic lattices
far above the percolation threshold have been found to exhibit log-periodic oscillations in the
effective exponent versus time \cite{StauSor}. These log-periodic oscillations stem from an
Arrhenius exponentiation of a discrete spectrum of trapping well depths, leading to
a discrete hierarchy of time scales. This mechanism is very similar to the previous one.
Similar results have been found also for two-dimensional systems under the same
conditions \cite{Kirsch}. Julia Draeger in Hamburg has also
found log-periodic oscillations above the two-dimen-
sional percolation tthreshold with topological bias in diffusion, emphasized
by Havlin and Bunde a decade ago.  The topological bias has not a
fixed direction like an electric field, but points along the local direction
of the shortest distance (more appropriate for hydrodynamic flow). 

This mechanism for generating log-periodic oscillations makes use of an
interplay between dynamics and quenched randomness leading to a regime where the dynamics
is highly intermittent. The presence of the discrete lattice and the mesh size is essential.
This can be generalized\,: when the property of a system depends on extreme fluctuations
which have a size or amplitude equal to a multiple of a characteristic mesh size, log-periodic
oscillations can be expected. This mechanism thus applies also to the Lifshitz singularities
in random systems\,: exact calculations in random harmonic one-dimensional chains exhibit 
a (log-)periodic factor in the Lifshitz  essential singularity occuring near the van Hove
band-edge in the density of state \cite{Nieuluckk}. Intuitively, in the language of 
harmonic chains \cite{Nieuluckk}, this essential singularity
stems from the fact that the existence of a mode with a frequency close to the 
maximal frequency demands a large region containing light masses only.
Similar intermittent amplification processes have also been studied 
in \cite{Vlad,Vlad2}. Log-periodicity found in the solutions of boolean delay equations
\cite{boolean,bool2} stems from a similar mechanism.

Let us also mention Ref.\cite{Stefancich} who report log-periodic oscillations in the
survival probability for particles at the interface between an accelerator mode and the
chaotic sea in the kicked rotor probelm. 
The log-periodicity seems to be reinforced by quantum effects. Its origin could be the
existence of a regular time scale, the kicking time, which plays the role
of a ultraviolet cut-off and the anomalous trapping/acceleration leads to an
exponentiation of this arithmetic (i.e. additive) discrete time scales.
Thus discrete scale invariance comes here from the same type of
exponentiation of a discrete translational invariance, as before.

\subsection{Cascade of ultra-violet instabilities\,: growth processes and rupture}

\subsubsection{Log-periodicity in the geometrical properties}

Numerical analysis of large diffusion-limited-aggregates have uncovered a
{\it discrete} scaling invariance in their inner structure, which can
be quantified by the introduction of a set of {\it complex} fractal
dimensions \cite{DLA}. The values of the complex fractal dimensions can 
be predicted {\it quantitatively} from a renormalization group approach using the
quasi-periodic mapping found in \cite{arneodo}. 

A theoretical investigation of a simplified model of DLA, the needle problem, which is
also of direct application to crack growths has been done to identify the underlying
physical mechanism \cite{needles}. Based on perturbative analysis and some exact results
from the hodograph method in the 2D conformal plane, we find that the two basic
ingredients leading to DSI are the short wavelength Mullins-Sekerka instability
{\footnote{The Mullins-Sekerka instability is nothing other than the ``lightning rod
effect'' well-known in electrostatics, according to which large curvature concentrate
the gradient of the potential field. Here, the growth velocity is proportional to the
gradient of the concentration field.}} and the strong screening of
competing needles. The basic simple picture that emerges is that {\it non-linear}
interactions between the unstable modes of the set of needles lead to a succession of
period doubling, the next sub-harmonic catching up and eventually screening the leading
unstable mode. The succession of these period doubling
 explains the existence of discrete scale invariance in these
systems. We thus think that short wavelength instabilities of the Mullins-Sekerka type
supplemented by a strong screening effect provides a general scenario for the {\it
spontaneous} formation of log-periodic structures. This scenario provides, in
addition, an explanation for the observation of a prefered scaling ratio
close to $2$.

Numerical simulations of the
needle problem, using various growth rules (DLA, angle screening,
$\eta$-model, crack approximation) on systems containing up to $5000$
needles confirm clearly the proposed scenario, as shown in figure 4.
The density of needles as a function of the distance to the base
presents clear evidence of log-periodic modulations of the leading algebraic
decay. Geological data on joints competing in their growth in a similar fashion also
exhibit approximately the log-periodic structure \cite{needles,Ouillon}.
\cite{Degraff,Pollard} present further data on joints which, in our eyes, exhibit
clearly log-periodicity, even if the authors were not aware of the concept. Various
previous investigation of the growth of arrays of cracks have shown the log-periodic
structures, even if the authors neither point it out nor explained the mechanism
\cite{Nemat1,Nemat2}.

What we learn by comparing these different systems, with various
growth rules, is that the spontaneous formation of DSI seems robust with respect to
significant modifications. The improvement of our understanding of DLA resides on the
identification of a spontaneous generation of an approximately discrete cascade of
Mullins-Sekerka instabilities from small scales to large scales. This discreteness
results from a cascade of mode selections by a nonlinear nonperturbative
coupling between modes of growth \cite{needles}. Let us mention that, in the early
eighties, Sadovskiy et al. have argued for the existence of a discrete hierarchy in
fracture and rock properties \cite{Sadovskiy}, with a prefered scaling ratio around
$3.5$. Borodich \cite{Borodich} discusses the use of parametric-homogeneous functions for
a parcimonious mathematical representation of structure presenting a hierarchy of
log-periodicities.

In their theoretical analysis, Ball and Blumenfeld \cite{Ball} predicted logarithmic oscillations
in quasi-static crack growth, probably one of the very first example of 
such oscillations in non-tree structures.
First, they coarse-grained a quasi-static growing crack as a
wedge and  found the behavior of the stress field around it. Then, they showed through a
linear stability analysis that there is an instabitily to growth of
branches where the  stress is locally high. In this, it is conceptually similar to the DLA
instability with respect to a locally high gradient of the field near the interface. Then 
they argued that because of this instability the dynamics couple to sub-dominant terms
in  the stress field to generate branches. So looking into what are the
strongest subdominant terms, they could identify the behaviour of the branching. These
terms happened to have the functional form 
$$
S_n = A_n ~r^{a_n + ib_n}~,
$$
where $A_n$, $a_n(<0)$, and $b_n$ are numbers and $S_n$ is the stress due to 
the contribution of the $n$-th term in the expansion of the stress field.
Their argument was that, since these terms exhibit logarithmic oscillations,
then so will the branching in the cracking pattern, resulting in  a growing
dendrite with logarithmically periodic branches.

Let us finally mention that the distribution of avalanche strengths in the train model,
a variant of the Block-Spring Burridge-Knopoff model of earthquakes pulled at one
extremety, is a powerlaw times a log-periodic function \cite{Elmer}. This reflects
the spontaneous formation of a discrete hierarchy. In this model,  
the prefered scaling ratio in the length size of the avalanches is equal to  $2$. 
The moving blocks obey a discrete wave equation with attenuation at large velocities.
A non-local cellular automata model retrieves the results and is defined so as to mimick
the behavior of the train model. The main point is that the active sites of the cellular
automata obey a discrete Laplace equation. In the overdamped approximation valid for large
attenuation, the train model is also in this regime.
The discrete hierarchy of the sliding events reflects the instabilities of the growth processes
controlled by the Laplace equation

\subsubsection{Log-periodicity in time}

A growing body of evidences indicate that the log-periodic oscillations appear in the
time dependence of the energy release on the approach of impending rupture in
laboratory experiments \cite{Anifrani}, numerical simulations \cite{Sahimi}
and earthquakes \cite{SorSam,SSS1,SSS2,Kobe,Varnes,Bow1,Smith}.  It is thought that
a similar type of cascade, from progressive damage at small scale to
coalescence and unstable growth, is controlling the appearence of log-periodicity.
The typical time-to-failure formula used in these works is
\be
E \sim (t_r - t)^m \biggl[1 + C cos\biggl(2 \pi
\frac{\log (t_r-t)}{\log \lambda} + \Psi \biggl) \biggl]~~~~,
\label{AE}
\ee
where $E$ is the energy released or some other variable quantifying the on-going damage,
$t_r$ is the time of rupture, $m$ is a critical exponent, and $\Psi$ is a phase in the
cosine that can be get rid of by a change of time units. It has been found that the
log-periodic oscillations enable a much better reliability of the prediction due to
``lock-in'' of the fit on the oscillating structure. Physically, the oscillations
contain information on $t_r$ and thus help significantly in its determination. A link
between log-periodicity in space and in time is given in \cite{SOCTT}. A typical fit
by expression (\ref{AE}) to acoustic emission data is presented in figure 5.

\subsection{Cascade of structure in hydrodynamics}

Moffatt \cite{Moffatt} has studies similarity solutions for the flow of a viscous fluid
near a sharp corner between two plates on which a variety of boundary conditions are
imposed. For this, one has to solve the biharmonic Stokes equation for the stream
function which admits separable solutions in plane polar coordinates $(r,\theta)$\,: 
$\Psi = r^{\alpha} ~f_{\alpha}(\theta)$. For angles between the two plates less
than $146$ degrees, the exponent $\alpha$ has been shown to be necessarily complex
\cite{DeanMontagnon}. Moffatt has shown that this result can be interpreted as implying
the existence of an infinite sequence of eddies near the corner. It is interesting
that viscosity, usually a damping mechanism, is here responsible for the generation of
a geometrical progression of eddies. The damping has the effect, however, to give a 
large ratio (typically greater than $300$) of the intensities between successive eddies.

Recent experiments and analysis both numerical and analytical of droplet fission shows the 
existence of iterated instabilities that develop a discrete scale invariance. The
reason for self-similarity is that, near breakoff, the droplet radius becomes much
smaller than any other length scale, so that the shape of the interface becomes
independent of these scales. Numerical simulations of the corresponding
hydrodynamic equations with a weak noise source show that necks and blobs form 
repeatedly on smaller and smaller scales as the interface breaks \cite{Brenner}.
These instability cascades are only observed for fluids with a viscosity greater
than $1~P$. The mechanism behind this cascade of instabilities is that, immediately
before a neck forms, the thinnest section of the interface is well approximated by
the similarity solution. The nonsteady singularity results from repeated instabilities
of the similarity solution due to thermal capillary waves.

\subsection{Deterministic dynamical systems}

\subsubsection{Cascades of sub-harmonic bifurcations in the transition to chaos}

An area where log-periodic structures should be expected is
low-dimensional dynamical systems exhibiting the Feigenbaum
sequence of subharmonic bifurcations to chaos \cite{Feigen}. Indeed, this route to
chaos  can be understood from an asymptotically exact
{\it discrete} renormalization group with a universal scaling factor. The existence
of this prefered scaling ratio should thus lead to 
complex exponents and log-periodic oscillations around the main scaling
as the dynamics converges to the invariant Cantor set measure at criticality.
It was noticed quite early \cite{Derrida1} that the length of the stable period
diverges as a power law with log-periodic modulations as the control parameter
approaches the transition to chaos.
Argoul et al. \cite{Argoul} have studied the transitions to chaos in the presence of an
external periodic field and show figures exhibiting very clearly that the Lyapunov
exponent has a power dependence with log-periodic oscillations as a function of the
amplitude of the external field. Similarly, the topological entropy at the onset of
pruning in generalized Baker transformations is a power law function of the distance to
the onset of pruning with log-periodic oscillations \cite{Vollmer}. The oscillations
are due to the self-similar structure of the Cantor set forming the attractor
\cite{Derrida1}.

\subsubsection{Two-coupled anharmonic oscillators}

Consider the Hamiltonian
\be
H = {p^2 \over 2} + {\alpha \over 4} (q^2 -b^2)^2 + {P^2 \over 2} + {1 \over 2} \Omega^2 ~u^2
+ {1 \over 2} g~u~(q^2-b^2)~,
\ee
where $(P,u)$ and $(p,q)$ are the two conjugate pairs of (momentum, displacement) for the 
two degrees of freedom $u$ and $q$. The Hamilton-Jacobi equation of motion are
\ba
{d^2 q \over dt^2} + \alpha~q~(q^2 - b^2) + g~u~q &=& 0\\
{d^2 u \over dt^2} + \Omega^2~u + {1 \over 2} g~(q^2 - b^2)  &=& 0~.
\ea
Suppose that $q$ is observable while $u$ is not. One can see the problem as the motion
of a single particle in the time dependent potential
\be
V_{\rm eff} = {\alpha \over 4}(q^2 -b^2)^2 + {1 \over 2} \Omega^2~ u^2 + 
{1 \over 2} g~u~(q^2 -b^2)~,
\ee
where $u(t)$ is a time-dependent parameter responding to the particle position $q(t)$.
The quartic potention $V_{\rm eff}$ exhibits two wells and the particle oscillates 
between in a chaotic fashion for sufficiently large coupling $g$. 
West and Fan \cite{Westfan} have shown that this system is chaotic with positive
Lyapunov exponent for sufficiently large coupling $g$. More interestingly, the spectrum
of $q(t)$, equal to the Fourier transform of the correlation function $\langle q(t) q(t+\tau)
\rangle$, is found to be a power law $\sim k^{-2}$ with log-periodique modulations. These
modulations are best seen in the so-called ``rescaled range analysis'' introduced by 
Hurst \cite{Hurst}, which emphasizes the extremal fluctuations of the signal and thus
the observation of the modulations. The log-periodic oscillations disappear 
progressively at large time scales, due probably to the fact that they constitute
a correction to scaling. Intuitively, one can propose the following qualitative
understanding. The time-dependence of $u$ makes the dynamics chaotic and also 
modulates the height of the barrier separating the two wells close to $\pm b$. For a given 
height and in the presence of chaotic behavior, one can expect a kind of Arrhenius
activated behavior controlling the time of passage from one well to the other. 
The log-periodicity would then result from a discrete possible sets of values of $u$
at which this cross-over occurs. Further analysis is needed to ascertain this
mechanism.

\subsubsection{Near-separatrix Hamiltonian chaotic dynamics}

In a series of papers, Zaslavsky et al. (see \cite{Zaslavsky} for a review) have 
introduced a renormalization group transform for Hamiltonian dynamical systems 
with a saddle-point (separatrix) presenting a logarithm dependence of the 
period of periodic orbits as a function of the distance in energy scale to the
saddle-point. The scale transformation performed in the renormalization group analysis
corresponds to change both the strength $\epsilon$ of the perturbation 
applied to the system and the distance $h$ to the saddle-point. In analogy
with usual critical points, $\epsilon$ plays the role of the scale $L$ at
which the systems is coarse-grained and $h$ is the distance to the critical point.
For the harmonically perturbed pendulum and for the motion of two
particles in a fourth-order polynomial potential,
the renormalization group involves a prefered scaling ratio $\lambda$
for the scale transformation (eq.3.17 and eq.4.14 in \cite{Zaslavsky}). 
In the language of this review, this corresponds to having a
discrete scale invariance. The consequence is that suitable observables
should have complex critical exponents and the signals should be decorated
by log-periodic signatures.


\subsection{Animals}

We have notice \cite{SS} that, in contrast 
to common lore, complex critical exponents should generally be expected in
the field theories that describe geometrical systems, because the latter are {\it non
unitary}. In particular, evidence of complex exponents in lattice animals, a simple
geometrical generalization of percolation has been presented \cite{SS}.
The model of lattice animals is the most natural generalization of the
percolation model, which itself is the prototype of disordered systems.
The animal problem is the statistics
of connected clusters on a lattice  \cite{LI,Stauffer} and thus also describes
unrooted branched polymers.
Using transfer matrix techniques, the
number of unrooted branched polymers of size $N$ is found to exhibit
 a correction to the main scaling with a complex exponent. 

Recall that
in the percolation problem, bonds are occupied with a probability $p$
and unoccupied with probability $1-p$. For a given configuration, connected
parts are called clusters. To study the statistics of one percolation
cluster, one can sum over all configurations for the bonds that do not belong
to this cluster nor to its perimeter. Since they are either occupied or
inoccupied, the sum over all configurations just gives a unit weight. Hence
in percolation, clusters are simply weighed with $p^{N_b}(1-p)^{N_p}$ where
$N_b$
is the number of bonds in the cluster, $N_p$ the number of bonds of the
perimeter. Now the animal problem is a more general model
where a cluster is weighed by $p^{N_b}q^{N_p}$ with general values of
$p,q$. By varying $p,q$ a critical point is met which is always in the
same universality class of so-called animals. Only when $p+q=1$
is this critical point in a different universality class, percolation, which
therefore can be considered as a tricritical point in the
animals parameter space. 

The result of the analysis of \cite{SS} is that the  number $T_N$
of unrooted branched polymers of size $N$ in the
plane is given by
\be
T_N \approx \left({1\over p_c}\right)^N\left[N^{\nu(2-2X_1)-3}+c
N^{\nu(2-2X_{2,R}-3)}\cos\left(2X_{2,I}\nu\log N +\phi\right)\right]~~~~,
\label{fff}
\ee
where $\nu$ is the radius of gyration exponent, $\nu\approx 0.64$. Recall
that the leading term in (\ref{fff})  is actually known
exactly $\nu(2-2X_1)-3=-1$ \cite{Parisi}. Hence we see log-periodic terms to appear in
the next to leading behaviour of $T_N$. Unfortunately, since conformal invariance is
broken, this argument does not allow us to make any predictions on the amplitude of
these terms, which might well be very small. The DLA problem is much more favorable in
that respect probably due to enhancement effects stemming from the long-range
interactions of Laplacian fields.

\subsection{Quenched disordered systems}

Renormalization group analysis of a variety of spin problems with long-ranged quenched
interactions have found complex critical exponents 
\cite{Aharony,spinglass,Khmel,Boya,Weinrib}. However, these authors have in general
remained shy as to the reality of their results. Indeed, one could argue that
uncontrolled approximations (present in all these works) rather than physics could be the
cause of the complex exponents. Derrida and Hilhorst have also found log-periodic
corrections to the critical behavior of 1D random field Ising model at low temperature
by analyzing products of random non-commutative matrices \cite{Hilh}.

With the qualitative understanding of the ultrametric structure of the energy landscape
of spin glasses \cite{Mezard,Rammal} in the mean field approximation, one could conjecture that
these above results could be the observable signature of the hierarchical structure of
energy states in frozen random systems. The problem is that, even if hierarchical, the
ultrametric structure is believed to be continuous and it is not clear what could
produce the discrete scale symmetry.
It is generally believed 
that such topology occurs more generally
in other complex systems  with highly degenerate, locally stable
states \cite{stein}. However, much works remain to be done to clarify
this problem. The additional presence of long-range interactions complicate the matter further.

A dynamical model describing transitions between states in a hierarchical system of
barriers modelling the energy landscape in the phase space of meanfield
spinglasses leads again to log-periodic corrections to the main $\log t$ behavior
\cite{SS}.

Let us also mention that $m$th critical Ising models ($m=1$ for the Gaussian model,
$m=2$ for Ising, $m=3$ for tricritical Ising, ...)  have a free energy exhibiting
log-periodic oscillations as a function of the control parameter for large $m$, 
a signature of the
geometrical cascade of multicritical points \cite{Las1,Las2,Ludwig,SS}.

\section{OTHER SYSTEMS}

\subsection{The bronchial tree}

It has been pointed out that the morphology of the bronchial
airway of the mammalian lung is roughly hierarchical leading to a log-log
plot of the average diameter of a branch of a mammalian lung (for human, dog,
rat and  hamster) as a function of the branch order which exhibits a full
S-oscillation (log-periodic)
decorating an average linear (power law) dependence. This fractal
structure has been argued to allow the organ to be more stable with respect
to disturbance \cite{lung,West,Schlesin,Deering}  but the physical mechanism underlying
its appearence is not understood.

\subsection{Turbulence}

Probably the first theoretical suggestion of the relevance of log-periodic oscillations
to physics has been put forward by Novikov to describe the influence of intermittency in
turbulent flows \cite{Novikov}. The idea is that the DSI could stem from the existence
of a prefered ratio in the cascade from large eddies to small ones. The existence of
log-periodic oscillations has not been convincingly demonstrated as they seem quite
elusive and sensitive to the global geometry of the flow and
recirculation \cite{Ansel,Frisch}. 
Shell models of turbulence, which have attracted recently a lot of interest
\cite{shell} construct explicitely a discrete scale invariant set of equations. In these
models, self-similar solutions of the cascade of the velocity field and energy in the
discrete log-space scale have been unravelled \cite{soliton}, whose scaling can be
related to the intermittent corrections to Kolmogorov scaling. We note that some of these
solutions rely on the discrete scaling shell structure and would disappear in the
continuous limit. However, the relevance of these discrete hierarchical models
and more generally of log-periodic oscillations have not been explored
systematically and their confirmation in turbulence remains open. 

Based on theoretical argument and experimental evidence, the author has conjectured
that structure functions of turbulent times series exhibit
log-periodic modulations decorating their power law dependence \cite{Sorturb}.
In order to provide ironclad experimental evidence, we stress the need
for novel methods of averaging and propose a 
a ``canonical'' averaging scheme, amounting to 
condition the average of structure factors 
of turbulent flows on local maxima of the energy dissipation rate or the energy gradient.
The strategy is to determine the scale $r_c$ at which
the dissipation rate is the largest in a given turn-over time series. This
specific scale $r_c$ translates into a specific ``phase'' in the logarithm
of the scale which, when used as the origin, allows one to phase up the
different measurements of a structure factor $S_p(r) = A_p (\bar\epsilon
r)^{p/3}$ in different turn-over time realizations. We expect, as in
Laplacian growth and in rupture, that the log-periodic
oscillations will be reinforced by this canonical averaging. 
Demonstrating unambiguously the presence of log-periodicity and
thus of discrete scale invariance (DSI) in turbulent time-series would
provide an important step towards a direct demonstration of the Kolmogorov
cascade or at least of its hierarchical imprint.

\subsection{Titius-Bode law}

Dubrulle and Graner \cite{Titius} have noticed
that the Titius-Bode law of planets distance to the sun $r_n = r_0 K^n$ with $K
\approx 1.7$ can be seen a discrete scale invariant law ($K$ then plays the role of
$\lambda$ in our notation). 
They show that all models that have been proposed to explain the Titius-
Bode law share the common ingredient of scale symmetry. Assuming a discrete symmetry 
breaking in the rotation invariance, they thus show that any such mechanism is
compatible with the Titius-Bode law. As a consequence, this law cannot a
priori be used to constraint the mechanism of planet formation and their
organization around the sun. What is however not understood is the 
physical mechanism, if any, at the basis of the breakdown of continuous to discrete
scale invariance embodied in the Titius-Bode law.

\subsection{Gravitational collapse and black hole formation}

Choptuik \cite{Blackhole} has recently shown that, in contrast to the general view, black 
holes of mass smaller than the 
Chandrasekkar limit could be formed and that, in the process of formation, the solutions
would oscillate periodically in the logarithm of the difference between time and
time of the formation of the singularity. This gravitational collapse is an example
of critical behavior,
describing how the mass $M$ of the black hole depends on the strength $p$ of the
initial conditions\,: $M \sim (p-p^*)^{\gamma}$ for $p>p^*$ and $0$ otherwise, where
$p^*$ is the threshold value. It has been shown that classical close-to-critical black
holes (obeying Einstein's equations)
coupled to a massless complex scalar field have a leading real exponent $\gamma_R$
and a subleading complex exponent \cite{Blackhole}, which would correspond to a
log-periodic spectrum of masses. Alternatively, the real and complex exponents control
the time development of the black hole instability which is also log-periodic in time,
corresponding to continuous phase oscillations of the field.

\subsection{Spinodal decomposition of binary mixtures in uniform shear flow}

Corberi et al. \cite{Corberi} have studied the phase separation kinetics of a binary
mixture subjected to an uniform shear flow quenched from a disordered to a homogeneous
ordered phase. They used the time-dependent Ginzburg-Landau equation with an 
external velocity term.
They find numerically that the typical sizes of domains in the directions parallel and orthogonal to 
the direction of the shear as well as the excess viscosity exhibit log-periodic 
oscillations as a function of time that decorate the scaling laws. They have discovered
this behavior from the numerical integration of the Langevin equation for the evolution
of the system. From a theoretical point of view, the authors do not have an explanation  
for the log-periodic behavior of the scaling functions (private communication). 
They interpret these log-periodic oscillations
as due to a growth mechanism where stretching and break-up of domains occurs
cyclically in $\ln t$. They have performed a one-loop perturbation approximation, which 
together with the fact that (1) the natural time variable is $\ln t$ and
(2) the scaling function is assumed periodic allows them to retrieve the log-periodicity
of their numerical simulations. This is a first step, but it does not provide a
mechanism for log-periodicity since it has been ``put by hand'' in the scaling ansatz.
It is particularly interesting that the alternative stretching and break-up lead to
this cascade of scales. It would be of interest to pinpoint the underlying physical
mechanism by while a prefered scale ratio is selected.

\subsection{Cosmic lacunarity}

The present observable distribution of galaxies in space provides a signature
of their formation and interaction. 
There is an undergoing controversy on the nature of the spatial distribution of galaxies 
and galaxy clusters in the universe with recent suggestions that scale invariance 
could apply \cite{galaxy}. Recently, evidence of ``cosmic lacunarity'' has been 
presented \cite{Provenzale}\,: the prefactor of the power law dependence of the 
$q$th moment of the spatial galaxy distribution oscillates periodically 
with the logarithm of the distance $r$
with the same periodicity for $q=2, 3$ and $4$. The corresponding scaling factor is
around $2^{1.6} \approx 3$. A similar log-periodic oscillation has been suggested
much earlier \cite{Vaucouleurs} with approximately the same scaling ratio.
F. Combes (private communication) cautions that all previously announced 
periodicities (not log-periodicity) have finally been proven wrong. There are many
bias in the data and we have access only to the luminosity, not directly to the masses.
Furthermore, F. Combes argues that, since the existence of a prefered scaling ratio needs
to ``nucleate'' from either a small scale (like in DLA where the small scale is the
size of the particles) or from a large scale (maybe like turbulence), it is not
obvious what should be these cut-off scales in the present case.

\subsection{Rate of escape from stable attractors}

Let us also mention the recently discovered log-periodic behavior of the rate of escape
from a stable attractor surrounded by an instable limit cycle as a function of the
strength of the white noise \cite{Maier}. This is an example where the rate of escape,
as calculated from a Fokker-Planck equation, is non-Arhenius.

\subsection{Interface crack tip stress singularity}

Complex singularities are also found in the divergence of the stress as a function
of the distance to the tip of a crack at the interface between two different elastic
media \cite{Will}. The standard $\sigma \sim r^{-{1 \over 2}}$ singularity, where
$r$ is the distance to the crack tip and $\sigma$ is the stress, is replaced by
\be
\sigma \sim r^{-{1 \over 2} + i \omega}~~~~ ,
\ee
where
\be
\omega = {1 \over 2 \pi} \log \biggl(
{{{\kappa_1 \over \mu_1} + {1 \over \mu_2}} \over
{{\kappa_2 \over \mu_2} + {1 \over \mu_1}}} \biggl)~~~~ .     
\ee
Subscript $1$ and $2$ refer to the material in $y>0$ and $y<0$,
respectively; $\kappa = 3 - 4 \nu$ for plane strain and
$\kappa = {3-\nu \over 1+\nu}$ for plane stress, $\nu$ is the
Poisson ratio and $\mu$ is the shear modulus. Interface cracks
have important practical applications since interfaces between
composite media are often the locii of damage nucleation leading to the
incipient rupture. The existence of this complex singularity
suggests that the mechanism of damage and rupture at interfaces could
be quite different from that in the bulk \cite{Rice}.
Following this work, a wealth of studies have followed (see \cite{Rice}
and \cite{Riceb} and references therein), but the
physical understanding of the appearance of a complex critical singularity
has remained elusive.
Since the solution shows that the two modes of deformations in tension and
shear (modes I and II) are intrinsically coupled for an interface crack in
constrast to what happens for a crack in an homogeneous medium,  one could hope
to identify the physical origin of the complex exponent in this coupling. Let us also
mention that the pressure distribution as a function of distance to the corner in the
Hertz problem of two different elastic spheres compressed against each other is also
described by a power law with complex exponent \cite{Johnson}.

See also section 5.3.1 on Log-periodicity in the geometrical properties, which refers
to the work of Ball and Blumenfeld \cite{Ball}.

\subsection{The Altes'wavelet and optimal time-scale product}

The Altes wavelet has a log-periodic structure, with a local frequency varying
hyperbolically. It has the remarkable property of minimizing the time-scale 
uncertainty, in the same sense as the Gaussian law minimizes the time-frequency
uncertainty. As will become clear below, it also has the nice property that 
differentiating it corresponds to dilate it by a fixed factor.

Consider a signal $U(t)$ with Fourier transform $U(\omega)$. Let us define
\be
E_n \equiv {1 \over 2\pi}~\int_{-\infty}^{+\infty} d\omega~|(j\omega)^n~U(\omega)|^2~
\ee
as the energy of the $n$th signal component. Then
\be
V_n(\omega) = {(j\omega)^n~U(\omega) \over \sqrt{E_n}}
\ee
is an energy-normalized signal component. The rational to study $V_n(\omega)$
is that it corresponds to the normalized spectrum of the $n$th derivative of the signal $U(t)$.
The analysis of the signal $U(t)$ through the different $n$ components $V_n(\omega)$
corresponds to analyse the different derivatives of the signal.

Let the mean-square bandwidth of the function
$(j\omega)^n~U(\omega)$ be
\be
W_n^2 \equiv {\int_{-\infty}^{+\infty} d\omega_1~\omega_1^2~
|(j\omega_1)^n~U(\omega_1)|^2~ \over
\int_{-\infty}^{+\infty} d\omega_2~|(j\omega_2)^n~U(\omega_2)|^2
}~.
\ee
It is easy to verify from the Schwarz inequality that \cite{Altes}
\be
W_n^2 \geq W_{n-1}^2~.
\label{aeqgqm}
\ee
This result means that the band-width increases upon differentiating the function $U(t)$.
Altes proposed to measure signal and filter complexity in terms of time-bandwidth
product $T~W$. Equation (\ref{aeqgqm}) says that increasing $n$ will result in a larger
$W$, and in many cases $T~W$ will also increase. The components $V_n(\omega)$ will thus
generally become more complicated and more difficult to analyse as $n$ increases.

The idea of Altes \cite{Altes} is to postulate a signal with constant $T~W$\,:
\be
T_n~W_n = T_{n-1}~W_{n-1}~.
\label{hqhjwjjks}
\ee
This signal must have a root-mean-square duration $T$ that decreases when the time 
function is differentiated. The solution is that $V_n(\omega)$ satisfies
\be
V_n(\omega) \propto V_{n-1}(\omega/k)~,~~~~~~~{\rm where}~~k>1~.
\label{jgkkghoro}
\ee
Indeed, from (\ref{jgkkghoro}), $W_n = k~W_{n-1}$ and $T_n = T_{n-1}/k$. The
complexity of the components is the same\,: the inevitable bandwidth increase
is compensated by a proportionate decrease in duration.

It follows from (\ref{jgkkghoro}) that 
\be
V_n(\omega) \propto V_{0}(\omega/k^n)~,
\label{jqqghoro}
\ee
leading to
\be
\omega^n~U(\omega) \propto U(\omega/k^n)~.
\label{jgkkghorqqqo}
\ee
Specifying the constant which can depend on $n$ leads to
\be
\omega^n~U(\omega) = C(n)~ U(\omega/k^n)~.
\label{qqqqkghorqqqo}
\ee
This functional equation is very similar to the renormalization group equation 
previously encountered. The difference stems essentially from the fact that the constant 
$C(n)$ can depend on $n$. In the present approach, instead of changing the scale,
we change the order $n$ of derivation of the signal, which corresponds to 
changing the weight $\omega^n$ of the components. This corresponds to putting more and
more weight to the high frequencies, and thus, decreasing $n$ is like a coarse-graining.
Then, the condition (\ref{hqhjwjjks}) of equal complexity for all components
provides the self-similar function equation (\ref{qqqqkghorqqqo}). 
The solution of this functional equation is \cite{Altes}
\be
U(\omega) = A~\omega^{\nu}~\exp \biggl( - \biggl[ {(\ln \omega)^2 \over 2~\ln k}\biggl]\biggl)~
\exp \biggl( j {2\pi c~\ln \omega \over \ln k}\biggl)~,
\label{jkgkgkglg}
\ee
where 
\be
\nu = Re \{ {C'(0) \over k} \},~~~~~~2\pi~c = Im \{C'(0)\}~,
\ee
$A$ is a constant and
\be
C(n) = k^{n \nu + {n^2 \over 2}}~e^{j~2\pi~nc}~.
\ee
The origin of the log-periodicity of the waveform given by (\ref{jkgkgkglg}) is clear\,:
it is the conjunction of the fact that derivative orders $n$ are integers and that
each additional order of the differentiation acts exponentially on the frequency spectrum 
as proportional to $\omega^n$. This example is thus again a conversion of 
discrete translation to discrete scaling by an exponentiation mechanism.

$U(t)$ has a waveform which is a function of the parameters $\nu, k, c$. The parameter $k$
determines the spectral width, a larger $k$ implies a greater bandwidth. $k$ also
determines how much stretched is caused when the Fourier transform is multiplied
by $(j\omega)^n$, i.e. under $n$ derivatives of $U(t)$. The parameter $c$
determines the slope of the spectral phase function and is related to the ``chirp''
that appears in the times waveform. A larger value of $c$ leads to more zero crossings.
It can be shown that the waveform (\ref{jkgkgkglg}) is optimum as having a 
lack of Doppler coupling and a maximum Doppler tolerance \cite{Altes}.
Note that, by construction, all the components $V_n(\omega)$ have the same ratio of 
center frequency to bandwidth. Remarkably, these waveforms (\ref{jkgkgkglg}) are
very similar to those employed by animal sonars, such as bats and dolphins.

\subsection{Orthonormal Compactly Supported Wavelets with Optimal Sobolev Regularity}

Ojanen \cite{Ojanen} has recently presented a new construction of an
orthonormal compactly supported wavelets with Sobolev regularity
exponent as high as possible among those mother wavelets with a fixed
support length and a fixed number of vanishing moments. The increased
regularity is obtained by optimizing the locations of the roots the
scaling filter has on the interval $[\pi/2, \pi]$. The obtained wavelets have
a discrete scale invariant structure, as is apparent from their examination. This
is also seen directly by their construction.

To construct orthonormal compactly supported wavelets, Ojanen uses as the
starting point the discrete dilation equation
\begin{equation}\label{e:dileqn}
    \phi(x) = \sqrt{2} \sum_{k \in Z} 
	c_k \phi(2x-k), \qquad x \in R, \qquad c_k \in R,
\end{equation}
for the scaling function or father function~$\phi$. A solution exists
in the sense of distributions when $\sum_k c_k=\sqrt2$ and it has
compact support if only finitely many of the filter coefficients $c_k$
are non-zero (in fact, if $c_k\not=0$ only for $k\in\{0,\dots,n\}$
then the support of~$\phi$ is contained in~$[0,n]$). The solution of
the dilation equation is given by
\begin{equation}\label{e:m0prod}
    \widehat\phi(\xi) = \prod_{j=1}^{\infty} m_0(2^{-j}\xi),
\end{equation}
where the trigonometric polynomial $m_0(\xi)$ is the scaling filter
of~$\phi$,
\begin{equation}\label{e:m0}
    m_0(\xi) = \frac{1}{\sqrt2} \sum_{k\in Z} c_k e^{-ik\xi}.
\end{equation}

The condition when the solution to~\ref{e:dileqn} yields wavelets can
be expressed in terms of the scaling filter $m_0(\xi)$.
If $m_0$ satisfies
\begin{equation}\label{e:ortho}
    |m_0(\xi)|^2 + |m_0(\pi+\xi)|^2 = 1, \qquad \xi \in R,
\end{equation}
$m_0(0)=1$, and a technical condition
called the Cohen criterion holds, then there exists a scaling function and
wavelet pair.

If $\phi$ is such that it gives rise to
a wavelet, the wavelet $\psi$ is obtained by
\be
    \psi(x) = \sum_{k \in Z}  (-1)^k c_{1-k} \phi(2x-k).
\ee
Note that then $\phi$ and $\psi$ have the same regularity
properties. 

The original construction of orthonormal compactly supported wavelets
by Daubechies \cite{Daubechies} considers a trigonometric polynomial
$m_0$ with at most $2N$ non-zero coefficients $c_0$, \dots,
$c_{2N-1}$. The polynomial $m_0$ is required to
satisfy~\ref{e:ortho} and to have a zero at $\pi$ of maximal
order~$N$ (the order of the zero is also the number of vanishing
moments the corresponding wavelet has). The resulting equations can be
solved and the constructed $m_0$ has no roots on $(-\pi,\pi)$, hence it
satisfies the Cohen criterion.

To construct smoother wavelets, Ojanen's approach is to reduce the order of
the zero $m_0$ has at $\pi$ and instead to introduce zeros on the
interval $[\pi/2,\pi]$. Keeping $N$ and the number of zeros on $[\pi/2,\pi]$ fixed,
Ojanen uses numerical optimization to choose the locations of the roots so
that the resulting wavelets are as smooth as possible. He is thus able to
construct wavelets that are more regular than those introduced 
in \cite{Daubechies,Volkmer,Rieusset}.

\subsection{Eigenfunctions of the Laplace transform}

Log-periodicity and complex exponents play a very important role in integral
equations of the type $g(\tau) = \int_0^{\infty} K(v\tau) p(v) dv$, with $0 \leq \tau
< \infty$ where the kernel $K$ has the property $\int_0^{\infty} |K(x)| x^{-1/2} dx <
\infty$. This class of equation includes the Laplace transform, the Fourier sine and
cosine transforms and many other integral equations of importance in physics. It is
notorious that the inversion problem of getting $p(v)$ from the measurement of
$g(\tau)$ is ill-conditioned. This can be seen to result from the form of the
eigenfunctions and eigenvalues of the Laplace transform and similar dilationally
invariant Fredholm integral equation \cite{Pike}. For instance, the eigenfunctions of
the Laplace transform, which form a complete orthogonal basis, are
$\phi_{\omega}^{+}(v) =  v^{-1/2} \cos(\omega \ln v - \theta_{\omega})$ and
$\phi_{\omega}^{-}(v) = - v^{-1/2} \sin (\omega \ln v - \theta_{\omega})$, where
$\theta_{\omega}$ is a function of $\omega$. The eigenvalues are exponentially
decreasing with $\omega$ and this controls the ill-conditioned nature of the Laplace
inversion. The log-periodicity of the eigenfunctions lead to an optimal sampling
determined by a generalized Shannon theorem which obeys a geometrical series
\cite{Laplace}.

\subsection{Life evolution}

A very intringuing study has been performed by Chaline et al
\cite{Chaline}, who have found that the fossil equine of North America,
the primates and the rodents have followed an evolution path puntuated by major events
that seem to follow a geometrical time series of the type (\ref{rerfdffghw}),
with a prefered time scale factor $\lambda \approx 1.7$. Using t-student statistics
and Monte-Carlo tests, Chaline et al. find a very high statistical significance.
The critical time $t_r$ is roughly the present for the equines, in agreement with the
extinction of this species in North America ten thousand years ago (which is very short
compared to geological time scales), about two millions
years in the future for the primates and between 10 to 60 millions years in the future
for the rodents. The main problem is to ascertain that the determination of the
times of the major events has been done independently from the mathematical analysis. Otherwise,
this leads to a well-known statistical bias and the statistical significance of these
results looses its meaning.

\section{APPLICATIONS}

\subsection{Identifying characteristic scales}

In our opinion, the main interest in identifying log-periodicity in data is the
characterization of the characteristic scales associated to it. Indeed, it must be
clear that the log-periodic corrections to scaling imply the existence of a hierarchy of
characteristic scales (in space or time). For instance, in the time-to-failure analysis
given by (\ref{AE}), the hierarchy of time scales is determined by the 
local positive maxima of the function $E$. They are given by
\be
t_r - t_n = \tau \lambda^{n \over 2}~, 
\label{rerfdffghw}
\ee
where $\tau \propto \exp (-{\log \lambda \over 2 \pi} tan^{-1}{2 \pi \over m \log
\lambda})$. The
spacing between successive values of $t_n$ approaches zero as $n$ becomes large and
$t_n$ converges to $t_r$. This hierarchy of scales $t_r - t_n$ are not universal but
depend upon the specific geometry and structure of the system. What is expected to be
universal are the ratios $\frac{t_r - t_{n+1}}{t_r - t_n} = \lambda^{\frac{1}{2}}$.  
From three successive observed values of $t_n$, say $t_n$, $t_{n+1}$ and $t_{n+2}$, we
have 
\be 
t_r = {t_{n+1}^2 - t_{n+2}t_n \over 2t_{n+1} - t_n - t_{n+2}}~ .
\ee
This relation applies the Shanks transformation to acceleration of 
convergence of series. In the case of an exact geometrical series, 
three terms are enough to
converge exactly to the asymptotic value $t_c$. 
Notice that  this relation is invariant with respect to an arbitrary translation 
in time. In addition, the next time $t_{n+3}$ is predicted from the first three ones by
\be
t_{n+3} = {t_{n+1}^2 + t_{n+2}^2 - t_n t_{n+2} - t_{n+1} t_{n+2} 
\over t_{n+1} - t_n}~ . 
\label{sdf}
\ee
These relations have been used in \cite{Newman,Kobe,Varnes}.
Physically, time or space scales give us access to additional information
and clues about the underlying processes and the existence of a hierarchy
of prefered scales, as in DSI, will tell us something about the underlying 
processes. This is lost in usual critical behavior in which all scales are 
treated as playing the same role.

\subsection{Time-to-failure analysis}

Another important application of log-periodicity is its use in making more robust and
precise time-to-failure analysis. We have already mentionned the importance
of log-periodicity for predictions
\cite{SorSam,SSS1,Kobe,Anifrani,crash1,crash2}. The derived time-to-failure analysis is
now being implemented for routine industrial testing in the space industry in Europe.
As already mentionned, t
he reason for this improvement is that a fit can ``lock-in'' on the oscillations which
contain the information on the time of failure and thus lead to a better prediction.

\subsection{Log-periodic antennas}
Let us mention the engineering application of antennas using log-periodic
electromagnetic antennas 
\cite{SmithHH,Baker,Dykaar,Smithh,Excell,Delyser,Gitin}. The DSI structure provides an
optimal compromise between maximizing bandwidth and radiation efficiency.

\subsection{Optical waveguides}

Graded-index optical waveguides with optimized index profiles can
support a family of weakly localized modes with algebraic tails with log-periodic
modulations in the evanescent field \cite{Hayata}. The log-periodic oscillations result
from an interplay between the critical nature of the modes and absorption (complex
index of refraction). This could have applications in techniques using evanescent
waves.

\section{OPEN PROBLEMS}

\subsection{Nonlinear map and multicriticality}
 In this brief review,
we have kept the analysis of the renormalization group at the level of a linear
expansion of the flow map. Taking into account the nonlinear structure of the flow map,
as for instance in (\ref{mappp}), 
may lead to an infinite set of singularities accumulating at the main critical point or
even to the whole axis of the control parameter being critical (in the chaos regime of
the flow map \cite{Derrida}). The question of the relevance of these regimes to nature
is still open (see \cite{SSS1} for a proposed
 application to earthquakes). Generalization to several
control parameters and multicritical points would be useful.

\subsection{Multilacunarity and quasi-log-periodicity}
We have seen that DSI embodies the concept of lacunarity. The set of complex exponents
or singularities has been called {\it multilacunarity} spectrum \cite{Fournier2}.
Generalizations with several different incommensurate log-frequencies would be of great
interest and seem to appear for instance in the DLA problem \cite{DLA}. Complex
multifractal dimension spectrum in the presence of disorder can be handled using
probabilistic versions of the renormalization group \cite{Falconer} and their
development and impact are just emerging. Note also that
the set of complex exponents provides a better characterization of the underlying
multiplicative process and could improve the conditionning of the inverse fractal
problem \cite{Barnsley}. The $q$-derivative is a natural tool to discuss homogeneous
functions with oscillatory amplitudes. It has recently been used to describe cascade
and multifractal models with continuous scale changes \cite{Erzan}. In Ref.\cite{Erzan2},
a kinetics is built upon the $q$-calculus of discrete dilatations and is shown to 
describe diffusion on a hierarchical lattice.

\subsection{Effect of disorder}

A very important practical question is the effect of disorder and the process of
averaging. Disorder is expected to scramble the phases of the log-periodic oscillations
and it is a priori not clear whether the log-periodic oscillations are robust. 
It turns out that small fluctuations around the log-periodic structure do not seem to
spoil DSI, as found in many examples quoted above. For instance, in the needle DLA
problem, intervals between needles were taken to fluctuate by a few percent without
altering significantly the log-periodic structure of the growth process \cite{needles}. 
DLA clusters themselves are formed under a very strong annealed noise, corresponding to
the random walk motion of the sticking particles. Nevertheless, clear evidence of
log-periodicity in the mass as a function of radius has been found \cite{DLA} giving
confidence in their robustness with respect to disorder. This has been further
substantiated by explicit calculations \cite{SS} showing generally that the complex
exponents are robust.

\subsection{Averaging\,: grand canonical versus canonical}

However, we must stress that disorder introduces a sensitive dependence of the phase in
the $\cos \log$ formula \,: different realizations have a different phase and averaging
will produce a ``destructive interference'' that makes vanish the log-periodic
oscillations. It is thus important to carry out analysis on each sample realization
separately, {\it without} averaging. For instance, in the DLA case, 350 clusters of
$10^6$ particles have each been analyzed one by one and an histogram of the main
log-frequencies has been constructed. Theoretically, preventing averaging is a problem
as one is usually able only to calculate quantities averaged over the different
realizations of the disorder. However, it must be stressed that the fact that
log-periodic oscillations are mainly present before averaging tells us that they are
{\it specific} fingerprints of the specific system one is looking at. This is obviously
a desirable property for prediction purposes in engineering and other
practical applications. An open problem however is to devise optimal tools to decipher
the log-periodic structures in highly noisy data, as is usually the case due to
limitation of sizes for instance.  We note also that going to very large systems will
in general progressively destroy the log-periodic structures as they are often
correction to scaling {\footnote{This is often due to the disorder which can be shown
to renormalize  the real part of the complex exponents so that they correspond to
sub-leading correction to the main scaling behavior \cite{SS,DLA}.}}. Random versions
of Cantor fractal sets have recently been shown to exhibit robust log-periodic
structures even when averaging \cite{Solis}.

Pazmandi et al. \cite{Pazmandi} have recently argued that the standard method of 
averaging carried out
in disordered systems introduces a spurious 
noise of relative amplitude proportional to the inverse
square root of the system size. This so-called ``grand canonical'' averaging can thus
destroy more subtle fluctuations in finite systems, controlled by a correlation length
exponent less than ${2 \over d}$ (${2 \over d}$ is the 
minimum value of the correlation length exponent that would not be hidden by 
the usual grand canonical averaging). Pazmandi et al. thus propose an 
alternative averaging procedure, the so-called ``canonical'' averaging, which 
consists in identifying, for each realization, the corresponding specific value 
of the critical control parameter $K_c^R$. The natural control parameter
then becomes $\Delta = (K-K_c^R)/K_c^R$ and the act of averaging can then be
performed for the samples with the same $\Delta$. We have used this procedure to
identify log-periodicity in the elastic energy $E$ prior to rupture in 
a dynamical model of rupture in heterogeneous media. Previous works have shown
that $E$ follows a power law $E \sim (t_r - t)^{-\alpha}$ as a function of the time
to failure  \cite{Vanneste}. Performing the usual (grand canonical) averaging
over twenty different realizations of the disorder provides very good evidence
of the power law but no evidence of log-periodicity. We have thus developed the
following alternative averaging procedure \cite{finitezeiz}. We constructed the
second derivative of $E$ with respect to time for each realization,
thinking of it as a kind of 
susceptibility. The time $t_r^R$ at which this second derivative is maximum 
has been identified and
this point has been used as the effective value of the time $t_r$ of rupture
for each realization. Then, the first derivative, giving the rate of 
energy released, is averaged over all samples with the same $(t_r^R - t)/t_r^R$.
The result is presented in figure 6 in a log-log scale. Four to five approximately
equidistant spikes (in log scale) are clearly visible. The log-period allows us to
identify a prefered scaling ratio $\lambda = 2.5 \pm 0.3$. 
It is probable that similar averaging procedures better tailored to get
rid of spurious fluctuations from realization to realization will play an increasing
role in the physics of disordered and turbulent media.

\subsection{Amplitude of log-periodicity}

Log-periodicity is found in spin systems in hierarchical lattices. However, the effect
is usually very small, typically $10^{-4}$ or less in relative amplitude. In contrast,
we have found it much stronger in rupture and growth process, typically $10^{-1}$ or
so in relative amplitude. The reason is not very well understood but might stem from
the strong amplification effects occurring in such Laplacian fields.

There is a situation where the factors controlling the log-periodic amplitudes 
are understood. This is the case for spin models in hierarchical networks.
In this case, the free energy of the high temperature phase is in general analytic close to $T_c$
in a non-vanishing angular sector in the complex plane. As a consequence,
the Fourier coefficients $c_n$ of the log-periodic amplitudes decay exponentially with 
their order $n$. In practice, the first coefficient $c_1$ is already given by the
exponential law and therefore very small (see eq.(28) in \cite{DIL}).
In many other cases, the coefficients $c_n$ decay more slowly, typically as
powers of $n$, such that the Fourier sum which controls the amplitude of the
log-periodicity does not possess a domain of analyticity.
This occurs for instance for the density of states of harmonic chains with random
masses, both for the Lifshitz singularities at the band edges or at the 
Halperin singularities (cluster resonances) in the case of binary mixtures
\cite{Luckbook}. There are other cases where the log-periodic amplitudes may be of both
types depending on what we look at\,: in a chain of quantum $XY$ spins with quasiperiodic
modulated couplings, the singularities of the density of states correspond to the first
type (exponential with domain of analyticity) while the singularities corresponding
to the magnetization and the susceptibility are of the second type (power decay and
no domain of analyticity). See \cite{LuckNiew} and compare figures 3 and 4 to figures 5 and 6.

\subsection{Where to look for log-periodicity}

From the point of view of non-unitary field theory, we should expect generically
the existence of complex exponents. However, there is no known recipe to tell us what
are the relevant observables that will have complex dimensions. Practically, this means
that it is not a priori obvious what measure must be made to identify log-periodicity.
In other words, it is important to look carefully at available data in all imaginable
angles to extract the useful information. Of course, one must always been aware of
statistical traps that noise can be taken for log-periodicity. Analysis must thus be
carried out with synthetic tests for the null hypothesis, bootstrap
approach, etc (see for instance \cite{Kobe,needles} for the application of statistical
tests and the bootstrap method in this context).

\subsection{Prefered scaling ratio around $2$?}

Another puzzling observation is the value of the prefered ratio $\lambda \approx 2$,
found for a wide variety of systems, such as in growth processes, rupture, earthquakes,
financial crashes. H. Saleur (private communication) has noticed that $2$ is in
fact the mean field value of $\lambda$ obtained by taking an Ising or Potts model
(with $Q$ states) on a hierarchical lattice in the limit of an infinite number of
neighbors. Consider a diamond lattice with $n$ bonds connected to the upper and lower
nodes  (the usual diamond lattice discussed above 
has $n=2$). The discrete renormalization group
equation connecting  $K=e^{\beta J}$, where $\beta$ is the inverse temperature and $J$
the coupling coefficient, from one generation to the next is the generalized
version of (\ref{mappp}) to $n$-bonds connectivity\,:
\be
K'(K) = [f(K)]^n = ({K^2 + Q - 1 \over 2K + Q - 2})^n~~.
\label{fixedpoint}
\ee
In the limit $n \to \infty$ where the number of coupled nodes increases without
bounds, we expect physically the ordered-disordered transition to occur at larger and
larger temperature, corresponding to a fixed point of (\ref{fixedpoint}) 
$K'(K^*) = K^* \to 1$. Expanding around $1$, we indeed find $K^* = 1 + {Q \over n}$
asymptotically. The linearization of the renormalization group map (\ref{fixedpoint})
gives $K' - K^* = \lambda (K-K^*)$ with $\lambda = n K^* {d \log f \over d K}|_{K^*}
\to 2$ in the limit $n \to \infty$. Can this argument be extrapolated to
out-of-equilibrium systems?

\subsection{Critical behavior and self-organized criticality}

Time-to-failure analysis of earthquakes seem at variance with the globally stationary
view point, for instance captured by the concept of self-organized criticality as
applied to plate tectonics \cite{SOCtec}. Recently, it has been shown \cite{SOCTT}
that a simple model of earthquakes on a pre-existing hierarchical fault structure
exhibits both self-organization
at large times in a stationary state with a power law Gutenberg-Richter
distribution of earthquake sizes. In the same token, the largest fault carries
irregular great earthquakes preceded by precursors developing over long time
scales and followed by aftershocks obeying the ${1 \over t}$ Omori's
law of the rate of seismicity after a large earthquake. 
The cumulative energy released by precursors follows a
time-to-failure power law with log-periodic structures, qualifying a
large event as an effective dynamical (depinning) critical point. Down the hierarchy,
smaller earthquakes exhibit the same phenomenology, albeit with increasing
irregularities. The study of the robustness of this scenario for other models and
situations is an open question.

\vskip 0.5cm
Acknowledgments: I acknowlege useful discussions with J.-M. Luck and A. Provenzale.
I wish to thank my collaborators J.-C. Anifrani (Bordeaux,
France), A. Arneodo (Bordeaux, France), J.-P. Bouchaud (Saclay, France), Y. Huang
(USC, Los Angeles), C. Le Floc'h (Bordeaux, France),
J.-F. Muzy (Bordeaux, France), W. I. Newman (UCLA, Los Angeles), G. Ouillon (Nice,
France), C. Sammis (USC, Los Angeles), B. Souillard
(Orsay, France), D. Stauffer (Kohn, Germany), 
U. Tsunogai (Tokyo, Japan), C. Vanneste (Nice, France), H. Wakita
(Tokyo, Japan), who participated on diverse parts of this work. A special mention
should be given to A. Johansen (Copenhagen, Denmark) and H. Saleur (USC, Los Angeles).
I am also grateful to A. Johansen for a critical reading of the manuscript.
This is publication 4901 of the Institute of Geophysics and Planetary Physics
at UCLA.

\pagebreak

FIGURE CAPTIONS:
\vskip 1cm

Figure 1: Construction of the triadic Cantor set\,: the discrete scale
invariant geometrical structure is built by a recursive process in which
the first step consists in dividing
the unit interval into three equal intervals of length ${1 \over 3}$ and in deleting the
central one. In the second step, the two remaining intervals of length ${1 \over 3}$ are
themselves divided into three equal intervals of length ${1 \over 9}$ and their central
intervals are deleted, thus keeping $4$ intervals of length ${1 \over 9}$, and so on.
\vskip 1cm

Figure 2: Construction of the hierarchical diamond lattice used in the Potts model.
This lattice is  obtained by starting with a bond at
magnification 1, replacing this bond by four bonds arranged in the
shape of a diamond at magnification 2, and so on. The spins are placed at
the sites. At a given magnification $2^p$, one sees $4^p$ bonds, and
thus ${2\over 3}(2+4^p)$ sites.

\vskip 1cm

Figure 3:  ${\langle x^2(t) \rangle \over t^{2\nu_R}}$, where $\nu_R = 
{\log(1-p)\over \log(u/v)}$ is the real part of (\ref{ertdff}) as
a function of $\ln t$. The averaging has been performed over different
realizations of the random walk (taken from \cite{BS}).

\vskip 1cm

Figure 4: a) Map of $5000$ needles which have grown according to the DLA rules from an
initial configuration where all the needles were approximately of the same length equal
to their average separation. We have used a periodic lattice and 
added a small random value (a few percent of the period) for their lateral position.
This configuration corresponds to the time when the largest needle has a length equal to one-third
of the size of the system.
b) The probability density function of the needle lengths shown in a) in where very clear log-periodic
oscillations decorate the power law behavior. (see \cite{needles}).

\vskip 1cm

Figure 5: Logarithm of the acoustic emission energy released as a function of the pressure (in bars) 
applied within a pressure tank made of matrix-fiber composite at 
the approach of rupture. The continuous
lines correspond to the best fit by expression (\ref{AE}) (see \cite{Anifrani}).

\vskip 1cm

Figure 6: Rate of elastic energy released as a function of the 
logarithm of the time to failure in the dynamical model of rupture with damage 
introduced in \cite{Vanneste}. The dots have been obtained using the ``canonical'' 
averaging procedure discussed in the text.


\begin{thebibliography}{100}

\bibitem{Mandel} B.B. Mandelbrot, The fractal geometry of nature (San Francisco\,:
 W.H. Freeman, 1982).

\bibitem{Edgar} G.A. Edgar, ed., Classics on fractals (Addison-Wesley Publishing
Company, Readings, Massachussets, 1993).

\bibitem{revue} D. Sornette,
Discrete scale invariance and complex dimensions, Physics Reports 297, 239-270 (1998).

\bibitem{Zababakhin} E.I. Zababakhin, Shock waves in layered systems, J. Exp. Theoret.
Phys. 49, 642-646 (1965).

\bibitem{BarZEL} G.I. Barenblatt and Ya. B. Zeldovich, Intermediate asymptotics in
mathematic physics, Russian Math. Surveys 26, 45-61 (1971); Self-similar solutions
as intermediate asymptotics, Ann. Rev. Fluid Mech. 4, 285-312 (1972).

\bibitem{Novikov} E.A. Novikov, Sov. Phys. Dokl. 11, 497-499 (1966) 
[Dokl.Akad.Nauk SSSR 168/6, 1279 (1966)];  E.A. Novikov,
The effect of intermittency on statistical characteristics of turbulence and scale
similarity of breakdown coefficients, Phys.Fluids A 2,
814-820 (1990).

\bibitem{Jona} Jona-Lasinio G., The renormalization group: a probabilistic view,
Nuovo Cimento 26B, 99 (1975)

\bibitem{Nauen} Nauenberg M., Scaling representations for critical
phenomena, J.Phys. A 8, 925 (1975).

\bibitem{Nieme} Niemeijer Th. and J.M.J. van Leeuwen, in Phase transitions and
critical phenomena, Vol.6, C. Domb and M.S. Green, eds. (Academic Press, London,
1976), p.425. 

\bibitem{SS} H. Saleur and D. Sornette, Complex exponents and log-periodic
corrections in frustrated systems, J.Phys.I France 6, 327-355 (1996).

\bibitem{Bak} P. Bak, How nature works\,: the science of self-organized criticality (New
York, NY, USA\,: Copernicus, 1996).

\bibitem{BerryLewis} M.V. Berry and Z.V. Lewis, On the Weierstrass-Mandelbrot
fractal function, Proc. R. Soc. Lond. A 370, 459-484 (1980).

\bibitem{Weierstrassmore} B.R. Hunt, 
The Hausdorff dimension of graphs of Weierstrass functions,
Proc. Am. Math. Soc. 126, 791-800 (1998); G. Landini,
Is periodontal breakdown a fractal process? Simulations using the
Weierstrass-Mandelbrot function, J. Periodontal Research 32, 300-307 (1997);
M. VandenBerg and M. Levitin,
Functions of Weierstrass type and spectral asymptotics for iterated sets, 
Quarterly J. Mathematics 47, 493-509 (1996); D.C. Sun and Z.Y. Wen, 
The Hausdorff dimension of graph of a class of Weierstrass functions, 
Progress in Natural Science 6, 547-553 (1996); D. Nicoletti, 
Properties of the Weierstrass function in the time and frequency domains, 
Chaos Solitions \& Fractals 5, 1-8 (1995); T.Y. Hu and K.S. Lau, Fractal
dimensions and singularities of the Weierstrass type functions, Trans. Am. Math. Soc. 
335, 649-665 (1993); J. Gerling and H.J. Schmidt, Self-similar drums and generalized
Weierstrass functions, Physica A 191, 536-539 (1992); J.A.C. Humphrey, C.A. Schuler and
B. Rubinsky, On the use of the Weierstrass-Mandelbrot function to describe the
fractal component of turbulent velocity, Fluid Dynamics Research 9, 81-95 (1992).

\bibitem{Wallace} D. J. Wallace, R. K. P. Zia, Phys. Lett. 48A (1974)
325.

\bibitem{DLA} D. Sornette, A. Johansen,  A. Arn\'eodo, J.-F. Muzy and H. Saleur,
Complex fractal dimensions describe the internal hierarchical structure of DLA,
Phys. Rev. Lett. 76, 251-254 (1996).

\bibitem{Cardy} J. Cardy, J. Phys. A17 (1984) L385.

\bibitem{Weinberg} S. Weinberg, What is Quantum Field Theory, and what did we think it
is? (preprint hep-th/9702027); The quantum theory of fields,  Cambridge ; New York :
 Cambridge University Press, 1995.

\bibitem{SorSam}  D. Sornette and C.G. Sammis,
 Complex critical exponents from renormalization group theory of earthquakes : 
Implications for earthquake predictions, J.Phys.I France 5, 607-619 (1995)

\bibitem{SSS1}  H. Saleur, C.G. Sammis and D. Sornette, 
"Renormalization group theory of earthquakes", 
Nonlinear Processes in Geophysics 3, No. 2, 102-109 (1996)

\bibitem{SSS2}  H. Saleur, C.G. Sammis and D. Sornette, Discrete scale invariance,
complex fractal dimensions and log-periodic corrections in earthquakes,
J.Geophys.Res. 101, 17661-17677 (1996).

\bibitem{Kobe} A. Johansen, D. Sornette, H. Wakita, U. Tsunogai, W.I. Newman
and H. Saleur, Discrete scaling in earthquake precursory phenomena : evidence in the Kobe
earthquake, Japan, J.Phys.I France 6, 1391-1402 (1996).

\bibitem{Anifrani} J.-C. Anifrani, C. Le Floc'h, D. Sornette and B.
Souillard,
 Universal Log-periodic correction to renormalization group scaling for rupture
stress prediction from acoustic emissions, J.Phys.I France 5, n°6, 631-638
(1995)\,; J.-C. Anifrani, A. Johansen, C. Le Floc'h , G. Ouillon, D. Sornette, C.
Vanneste and B. Souillard, New approaches for exploiting acoustic emission²,
Proceedings of the 6th European Conference on Non-Destructive Testing, 24-28 october
1994, Nice, Presentation N72.

\bibitem{crash1} D. Sornette, A. Johansen and J.-P. Bouchaud,
Stock market crashes, Precursors and Replicas, J.Phys.I France 6, 167-175  (1996).

\bibitem{Freund} Feigenbaum, J.A., and P.G.O. Freund, Discrete
scale invariance in stock markets before crashes, Int. J. Mod. Phys. 10, 
N27, 3737-3745 (1996).

\bibitem{crash2} D. Sornette and A. Johansen,
Large financial crashes, Physica A 245, N3-4, 411-422 (1997).

\bibitem{Fournier} D. Bessis, J.-D. Fournier, G. Servizi, G. Tourchetti and S.
Vaienti, Phys. Rev. A 36, 920 (1987)

\bibitem{Fournier2} J.-D. Fournier, G. Tourchetti and S. Vaienti, 
Singularity spectrum of generalized energy integrals, Phys.lett.A 140, 331
(1989)

\bibitem{Orlan} E. Orlandini, M.C. Tesi and G. Turchetti, 
Corrections to the scaling laws of integrated wavelets from singularities 
of Mellin transforms, Europhysics Lett. 21, 719-722 (1993).

\bibitem{Kapitulnik} A. Kapitulnik, A. Aharony, G. Deutscher and
D. Stauffer, Self-similarity and correlations in percolation,
J. Phys. A 16, L269-L274 (1983).

\bibitem{Vladmackey} M.O. Vlad and M.C. Mackey, Multiple logarithmic oscillations for
statistical fractals on ultrametric spaces with application to recycle flows in
hierarchical porous media, Physica Scripta 50, 615-623 (1994)

\bibitem{Bessis} Bessis D., J.S. Geronimo and P. Moussa, J.Physique-Lett. 44,
L977-L982 (1983)

\bibitem{Doucot} Dou\c cot B., W. Wang, J. Chaussy, B. Pannetier and R. Rammal,
Phys. Rev. Lett. 57, 1235-38 (1986).

\bibitem{DIL} B. Derrida, L. De Seze and C. Itzykson, Fractal structure of zeros
in hierarchical models, J. Stat. Phys. 33, 559 (1983);
B. Derrida, C. Itzykson and J.M. Luck, Commun.Math.Phys. 94, 115-132
(1984)

\bibitem{Meurice} Y. Meurice, G. Ordaz, V.G.J. Rodgers, Evidence for complex
subleading exponents from high-temperature  expansion of Dyson's Hierarchical Ising
model, Phys.Rev.Lett. 75, 4555 (1995)

\bibitem{Newman} W.I. Newman, D.L. Turcotte and A.M. Gabrielov,
Log-periodic behavior of a hierarchical failure model with applications to
precursory seismic activation, Phys.Rev. E 52, 4827-4835 (1995)

\bibitem{Stanley} B. Kutjnak-Urbanc, Stefano Zapperi,  S. Milo\v sevi\'{c} and H.
Eugene Stanley, Sandpile Model on Sierpinski Gasket Fractal, Phys.Rev. E 54,
272-277 (1996).

\bibitem{LuckNiew} J.-M. Luck and Th. M. Nieuwenhuizen, A soluble quasi-crystalline
magnetic model: the $XY$ quantum spin chain, Europhys. Lett. 2 (4), 257-266 (1986).

\bibitem{Karevski} D. Karevski and L. Turban, Log-periodic corrections to scaling\,:
exact results for aperiodic Ising quantum chains, J. Phys. A 29, 3461-3470 (1996).

\bibitem{Vallejos} R.O. Vallejos, R. S. Mendes, L. R. da Silva and C. Tsallis,
Connection between energy-spectrum self-similarity and specific heat log-periodicity,
http://xxx.lanl.gov/abs/cond-mat/9803265; R. O. Vallejos and C. Anteneodo, 
Thermodynamical fingerprints of fractal spectra, http://xxx.lanl.gov/abs/cond-mat/9803294

\bibitem{Olemskoi} K.A. Makarov, Asymptotic expansions for Fourier transform of singular
self-affine measures, J. Math. Analysis and Applications 187, 259-286 (1994); 
Quantum scattering on a Cantor bar, J. Math. Phys. 35, 1522-1531 (1994);
A.I. Olemskoi, Fractals in condensed matter, Physics Reviews 18, 
1-173 (1995) (Harwood Academic, Malaysia).

\bibitem{Pecknold} S. Pecknold, S. Lovejoy and D. Schertzer, The morphology and texture
of anisotropic multifractals using generalized scale invariance, in Stochastic models
in geosystems, eds. S.A. Molchanov and W.A. Woyczynski, IMA Volumes in Mathematics and its
Applications, 85, 269-312 (1996).

\bibitem{Veneziano} D. Veneziano, G. Moglen and R.L. Bras, Iterate random pulse processes 
and their spectral properties, Water Resources (submitted)

\bibitem{Wu} F.Y. Wu, The Potts model, Rev. Mod. Phys. 54, 235 (1982).

\bibitem{Flajolet} P. Flajolet and M. Golin, Exact asymptotics of divide-and-conquer
recurrences, in Automata, Languages and Programming, 20th International Colloquium,
ICALP 93 Proceedings, Edited by Lingas, A., Karlsson, R., Carlsson, S.
(Springer-Verlag, Germany, 1993), 137-149.

\bibitem{Delange} P. Flajolet, P. Grabner, P. Kirschenhofer, H. Prodinger and R.F.
Tichy, Mellin transforms and asymptotics\,: digital sums, Theoretical Computer
Sciences 123, 291-314 (1994).

\bibitem{Gutzwiller} M.C. Gutzwiller, Chaos in classical and quantum mechanics (Springer, New
York, 1990).

\bibitem{Whan} C.B. Whan, Hierarchical level-clustering in two-dimensional
harmonic oscillators, Phys. Rev. E 55, R3813 (1997).

\bibitem{fractionexp} A. Ya. Khinchin, Continued fractions (University of Chicago Press,
Chicago, 1964).

\bibitem{onetoone} A. Pandey, O. Bohigas and M.J. Giannoni, Level repulsion in the 
spectrum of two-dimensional harmonic oscillators, J. Phys. A 22, 4083-4088 (1989);
A. Pandey and R. Ramaswamy, Level spacings for harmonic-oscillator systems, 
Phys. Rev A 43, 4237-4243 (1991).

\bibitem{BS} Bernasconi J. and W.R. Schneider, J.Phys.A15, L729-L734 (1983)

\bibitem{Nieuluckk} Th. M. Nieuwenhuizen and J.-M. Luck, Lifshitz singularities
in the total and the wavenumber-dependent spectral density of random harmonic chains,
Physica A 145, 161-189 (1987); Lifshitz singularities in random harmonic chains\,:
periodic amplitudes near the band-edge and near special frequencies, 
J. Stat. Phys. 48, 393-424 (1987).

\bibitem{Vlad} M.O. Vlad, Linear versus nonlinear amplification - a generalization
of the Novikov-Montroll-Shlesinger-West cascade, Int.J.Mod.Phys. B 6, 417-435 (1992)

\bibitem{Vlad2} M.O. Vlad, A new stochastic renormalization approach to random
processes with very long memory - fractal time as a process with almost complete
connections, J.Phys.A 25, 749-753 (1992)

\bibitem{boolean} Dee D. and M. Ghil, SIAM J. Appl. Math., 44,
111-126 (1984)

\bibitem{bool2} Ghil M. and A. Mullhaupt, J. Stat. Phys. 41, 125-173 (1985)

\bibitem{StauSor} D. Stauffer and D. Sornette,
Log-periodic Oscillations for Biased Diffusion on 3D Random Lattice, Physica A 252, 271 (1998)
(http://xxx.lanl.gov/abs/cond-mat/9712085)

\bibitem{Kirsch} A. Kirsch, Phase transition in two-dimensional biased diffusion, in press
in Int.J.Mod.Phys.C (1998).

\bibitem{Stefancich} M. Stefancich, P. Allegrini, L. Bonci, P. Grigolini and B.J. West,
Anomalous diffusion and ballistic peaks, a quantum perspective, Phys. Rev. E (preprint)
marco@soliton.phys.unt.edu

\bibitem{arneodo} A. Arneodo, F. Argoul, E. Bacry and
J. F. Muzy, Phys. Rev. Lett. 68 (1992) 3456; A. Arneodo, F. Argoul, J. F. Muzy and M.
Tabard, Phys. Lett. A 171 (1992) 31; A. Arneodo, F. Argoul, J. F. Muzy, M. Tabard, and
E. Bacry, Fractals, 1 (1993) 629.

\bibitem{needles} Y. Huang, G. Ouillon, H. Saleur and D. Sornette, Spontaneous
generation of discrete scale invariance in growth models,  
Physical Review E 55, 6433-6447 (1997).

\bibitem{Ouillon} G. Ouillon, D. Sornette, A. Genter and C. Castaing,
The imaginary part of the joint spacing distribution, J. Phys. I France 6,
1127-1139 (1996)

\bibitem{Degraff}	DeGraff, J.M., and A. Aydin : Effect of thermal regime on
	growth increment and spacing of contraction joints in basaltic lava,
	JGR, 98, B4, 6411-6430 (1993)

\bibitem{Pollard} Pollard, D.D., and A. Aydin : Progress in understanding jointing over
	the past century, Geol. Soc. Am. Bull., 100, 1181-1204 (1988)

\bibitem{Nemat1} S. Nemat-Nasser, L.M. Keer and K.S. Parihar, Unstable growth of
thermally induced interacting cracks in brittle solids, Earthquake research and
engineering lab., Technical report No.77-9-2, Department of Civil Engineering,
Northwestern University, Sept. 1977,International Journal of Solids and Structures
14, 409 (1978) 

\bibitem{Nemat2} S. Nemat-Nasser and A.Oranratnachai, Minimum spacing of
thermally induced cracks in brittle solids, Transactions of the ASME 101, 34-39 (1979)

\bibitem{Sadovskiy} M. A. Sadovskiy, The natural piecework of rocks, 
Dokl. Akad. Nauk SSSR, 247, 4, 829, 1979; M. A. Sadovskiy, L. G. Bolkhovitinov and
V. F. Pisarenko, Discrete properties of rock, Izv. Akad. Nauk SSSR, Fizika Zemli, 12,
3-19, 1982; M.A. Sadovskiy and V. F. Pisarenko, Several concepts of the seismic
process, Inst. Fiz. Zemli, Akad. Nauk SSSR, 2, 1982; 
M.A. Sadovskiy, V. F. Pisarenko and V. N. Rodionov, From seismology to geomechanics,
Vestn. Akad. Nauk SSSR, 1, 82-88, 1983; M.A. Sadovskiy, T.V. Golubeva, V.F. Pisarenko
and M.G. Shnirman, Characteristic dimensions of rock and hierarchical properties of
seismicity; Izvestiya, Earth Physics 20, 87-96, 1984.

\bibitem{Borodich} F.M. Borodich, 
Some applications of the fractal parametric--homogeneous functions,
Fractals 2, 311--314 (1994); Parametric--Homogeneous Functions, Similarity, and Fractal 
Function Graphs, Technical Report TR/MAT/FMB/95-35, Glasgow 
Caledonian University, Glasgow, 1--55 (1995); 
Applications of parametric--homogeneous functions 
to problems of interaction between fractal surfaces.
ZAMM 76 (S5), 61--62 (1996); Renormalization schemes for earthquake prediction,
Geophysical J. International 131(1), 171--178 (1997); 
Parametric-homogeneity and self-similar phenomena,
Nonlinear Analysis 30(1), 409--418 (1997); 
Some fractal models of fracture, J. Mech. 
Physics of Solids 45, 239-259 (1997); 
Parametric--homogeneity and non--classical self--similarity. 
I. Mathematical background (to be published Acta Mechanica);
Parametric--homogeneity and non--classical self--similarity.
II. Some applications (to be published Acta Mechanica)


\bibitem{Sahimi} M. Sahimi and S. Arbabi, Scaling laws for fracture of heterogeneous
materials and rock, Physical Review Letters 77, 3689-3692 (1996).

\bibitem{Ball} R.C. Ball and R. Blumenfeld, Universal scaling of the stress
field at the vicinity of a wedge crack in two dimensions and oscillatory self-similar
corrections to scaling, Phys. Rev. Lett. 65, 1784-1787 (1990);
 - Reply, Phys. Rev. Lett. 68, 2254-2254 (1992).
 
\bibitem{Elmer} F.-J. Elmer, Self-organized criticality with complex scaling exponents
in the train model, Phys. Rev. E 56, R6225-28 (1997).

\bibitem{Varnes} D.J. Varnes and C.G. Bufe, The cyclic and fractal
seismic series preceding and $m_b$ $4.8$ earthquake on 1980 February
14 near the Virgin Islands, Geophys. J. Int.124, 149-158 (1996)

\bibitem{Bow1} Bowman, D.D., and C.G. Sammis, An Observational Determination of the
Critical Region Before the 1983 M=6.7 Coalinga Earthqake (abst.), EOS
Trans. Am. Geophys. U., 77, page F486, 1996.

\bibitem{Smith} Smith, S.W., and C.G. Sammis, Discrete Hierarchic Cellular Model for
Foreshocks (abst.), EOS Trans. Am. Geophys. U., 77, page F480, 1996.

\bibitem{SOCTT} Y. Huang, H. Saleur, C. G. Sammis, D. Sornette,
Precursors, aftershocks, criticality and self-organized criticality, 
Europhysics Letters 41, 43-48 (1998).

\bibitem{Moffatt} H.K. Moffatt, Viscous and resistive eddies near a sharp corner, 
J. Fluid Mech. 18, 1-18 (1964).

\bibitem{DeanMontagnon} W.R. Dean and P.E. Montagnon, Proc. Camb. Phil. Soc. 45, 389 (1949).

\bibitem{Brenner} M.P. Brenner, X.D. Shi and S.R. Nagel, Iterated instabilities during
droplet fission, Phys. Rev. Lett. 73, 3391-3394  (1994); 
X.D. Shi, M.P. Brenner and S.R. Nagel, A cascade of structure in a drop falling from a 
faucet, Science 265, 219-222 (1994).

\bibitem{Feigen} M.J. Feigenbaum, J. Stat.
Phys. 19, 25 (1978); 21, 669 (1979); P. Coullet and C. Tresser, J. Phys. Coll. 39, C5
(1978); C.R. Acad. Sci. 287, 577 (1978); P. Collet and J.P. Eckmann, Iterated maps of the
interval and dynamical systems (Birkhauser, Boston, 1980).

\bibitem{Derrida1} B. Derrida, Propri\'et\'es universelles de certains syst\`emes
discrets dans le temps, J. Physique Colloques C5, 49 (1978);
B. Derrida, A. Gervois and Y. Pomeau, Ann. Inst. H. Poincar\'e AXXIX,
305 (1978).; Universal metric properties of bifurcations of
endomorphisms, J. Phys. A 12, 269 (1979).

\bibitem{Argoul} F. Argoul, A. Arn\'eodo, P. Collet and A. Lesne, Transitions to chaos
in the presence of an external periodic field\,: cross-over effect in the measure of
critical exponents, Europhysics Letters 3, 643-651 (1987).

\bibitem{Vollmer} J. Vollmer and W. Breymann, Scaling behaviour in Lorenz-like maps at
the onset of pruning,  Europhysics Letters 27, 23-28 (1994).

\bibitem{Westfan} B.J. West and X. Fan, Chaos, noise and complex fractal dimensions,
Fractals 1, 21-28 (1993).

\bibitem{Hurst} H.E. Hurst, Trans. Am. Soc. Civ. Eng. 116, 770 (1951).

\bibitem{Zaslavsky} L. Kuznetsov and G.M. Zaslavsky,
Hidden renormalization group for the near-separatrix Hamiltonian dynamics,
Phys. Rep. 288, 457-485 (1997).

\bibitem{LI} T. C. Lubensky, J. Isaacson, Phys. Rev. Lett. 41 (1978) 829.

\bibitem{Stauffer} D. Stauffer and A. Aharony, ``Introduction to percolation theory'',
second edition, Taylor and Francis (1992).

\bibitem{Parisi} G. Parisi, N. Sourlas, Phys. Rev. Lett. 46 (1981) 453.

\bibitem{Aharony} Aharony A., Critical properties
of random and constrained dipolar magnets, Phys. Rev. B 12, 1049-1056 (1975)

\bibitem{spinglass} Chen J.-H.and T.C. Lubensky, Mean field and
$\epsilon$-expansion study of spin
glasses, Phys. Rev. B 16, 2106-2114 (1977)

\bibitem{Khmel} D.E. Khmelnitskii, Impurity effect on the phase transition at T=0 in
magnets. Critical oscillations in corrections to the scaling laws, Phys.
Lett. A67,59-60 (1978).

\bibitem{Boya} Boyanovsky D. and J.L. Cardy, Critical behavior of $m$-component
magnets with correlated impurities, Phys. Rev. B 26, 154-170 (1982)

\bibitem{Weinrib} A. Weinrib and B.I. Halperin, Critical phenomena in systems
with long-range-correlated quenched disorder, Phys. Rev. B 27, 413-427 (1983)

\bibitem{Hilh} B. Derrida and H. Hilhorst, J. Phys. A 16, 2641-2654 (1983).

\bibitem{Mezard} M. M\'ezard, Parisi G. and Virasoro M.A., ``Spin glass theory and
beyond'', World Scientific, Singapore (1987).

\bibitem{Rammal} R. Rammal, G. Toulouse and Virasoro M.A., Rev.Mod.Phys.58, 765
(1986).

\bibitem{stein} D.L. Stein, in ``Chance and Matter'', J. Souletie, J.
Vannimenus and R. Stora eds., North Holland, Amsterdam (1987).

\bibitem{Las1} M. L\"assig, New hierarchies of multicriticality in
two-dimensional field theory, Phys. Lett. B 278, 439-442 (1992) 

\bibitem{Las2} M. L\"assig, Multiple crossover phenomena and scale hopping in two
dimensions, Nucl. Phys. B 380, 601-618 (1992)

\bibitem{Ludwig} A.W.W. Ludwig, Infinite hierarchies of exponents in a diluted
ferromagnet and their interpretation, Nucl. Phys. B330, 639-680 (1990).

\bibitem{lung} B.J. West, Int.J.Mod.Phys.B 4, 1629 (1990)

\bibitem{West} B.J. West, Annals of Biomedical Engineering, 18, 135 (1990)

\bibitem{Schlesin} Schlesinger M.F. and B.J. West, Phys.Rev.Lett. 67, 2106-08 (1991)

\bibitem{Deering} B.J. West and W. Deering, Phys. Rep. 246, 1-100 (1994)

\bibitem{Ansel}  Anselmet F., Y. Gagne, E.J. Hopfinger
and R.A. Antonia, High-order velocity structure functions in turbulent
shear flows, J. Fluid Mech.140, 63 (1984)

\bibitem{Frisch} U. Frisch, Turbulence, the legacy of A.N. Kolmogorov (Cambridge
University Press, 1995), p.130-131.

\bibitem{shell} M.Yamada and K. Okhitani, Phys.Rev.Lett. 60, 983 (1988); K.
Okhitani and M. Yamada, Prog.Theor.Phys. 81, 329 (1989); M.H. Jensen, G.
Paladin and A. Vulpiani, Phys.Rev.A 43, 7798 (1991),

\bibitem{soliton} T.Nakano, Prog.Theor.Phys.
79, 569 (1988) ; T. Dombre and J.-L. Gilson, preprint oct.1995

\bibitem{Sorturb} D. Sornette, Discrete scale invariance in turbulence?
in the Proceedings of the Seventh European
Turbulence Conference (ETC-7), June 30-July 3 (1998) (Published by Kluwer, U.
Frisch, editor) (http://xxx.lanl.gov/abs/cond-mat/9802121)

\bibitem{Titius} F. Graner and B. Dubrulle, Titius-Bode laws in the solar system\,:
I. scale invariance explains everything, Astronomy and Astrophysics 282, 262-268
(1994); II. Build your own law from disk models, 
 Astronomy and Astrophysics 282, 269-276 (1994).

\bibitem{Blackhole} M.W. Choptuik, Universality and scaling in gravitational collapse of
a massless scalar field, Phys. Rev. Lett. 70, 9-12 (1993); A.M. Abrahams and C.R.
Evans, Critical behavior and scaling in vacuum axisymmetric gravitational collapse, 
Phys. Rev. Lett. 70, 2980-2983 (1993);
Critical phenomena and relativistic gravitational collapse, General Relativity
and Gravitation 26, 379-384 (1994);
E.W. Hirschmann and D.M. Eardley, Critical exponents and stability
at the black hole threshold for a complex scalar field, Phys. Rev. D 52,5850-5856
(1995).

\bibitem{Corberi} F. Corberi, G. Gonnella and A. Lamura, 
Spinodal decomposition of binary mixtures in uniform shear flow, preprint 1998.
(gonnella@bari.infn.it)

\bibitem{galaxy} M. Montuori, et al., Europhys. Lett. 39, 103 (1997);
P.H. Coleman and L. Pietronero, Phys. Reports 213, 311 (1992);
M. Lachi\`eze-Rey, Scale invariance in the cosmic matter distribution ? in 
B. Dubrulle, F. Graner and D. Sornette, eds., Scale invariance and beyond 
(EDP Sciences and Springer, Berlin, 1997).

\bibitem{Provenzale} A. Provenzale, E.A. Spiegel and R. Thieberger, CHAOS 7, 82 (1997).

\bibitem{Vaucouleurs} G. de Vaucouleurs, Science 167, 1203 (1970).

\bibitem{Maier} R.S. Maier and D.L. Stein, Oscillatory behavior of the rate of escape
through an unstable limit cycle, Phys. Rev. Lett. 777, 4860-4863 (1996).

\bibitem{Will} M.L. Williams, Bull. Seismol. Soc. Am. 49, 199-204
(1959).

\bibitem{Rice} J.R. Rice, Transactions of the ASME, 55, 98-103 (1988)

\bibitem{Riceb} J.R. Rice, Z. Suo and J.-S. Wang, Mechanics and
thermodynamics of brittle interface failure in bimaterial systems,
in Metal-Ceramic Interfaces (eds. M. Ruhle, A.G. Evans, M.F.
Ashby ad J.P. Hirth), Acta-Scripta Metallurgica Proceedings Series,
vol. 4 (Pergamon Press, 1990), pp. 269-294.

\bibitem{Johnson} K.L. Johnson, Contact mechanics (Cambridge University Press,
Cambridge, UK, 1985), p.108; F.M. Borodich, The Hertz frictional problem contact
between nonlinear elastic anisotropic bodies (the similarity approach), Int. J.
Solids and Structures 30, 1513-1526 (1993); Fractal roughness in contact problems,
PMM Journal of Applied Mathematics and Mechanics 56, 681-690 (1992);
Similarity properties of discrete contact between a fractal punch and an elastic 
medium, Comptes Rendus Acad. Sci. Paris II, V316, 281-286 (1993).

\bibitem{Altes} R.A. Altes, Sonar for generalized target description and its
similarity to animal echolocation systems, J. Acoust. Soc. Am. 59, 97-105 (1976).

\bibitem{Ojanen} H. Ojanen,  Orthonormal Compactly Supported Wavelets with 
Optimal Sobolev Regularity, preprint math.CA/9807089.

\bibitem{Daubechies} I. Daubechies, Comm. Pure Appl. Math. 41, 909-996 (1988). 

\bibitem{Volkmer} H. Volkmer, Asymptotic regularity of compactly-supported wavelets, 
SIAM J. Math. Anal. 26, 1075-1087 (1995).

\bibitem{Rieusset} P.G. Lemari\'e-Rieusset and E. Zahrouni, More regular wavelets, 
Appl. Comput. Harmon. Anal. 5, 92-105 (1998).

\bibitem{Hayata} K. Hayata and M. Koshiba, Algebraically decaying modes in many
dimensions\,: application of Fubini's identity to guided-wave optics, Optical Review
2, 331-333 (1995).

\bibitem{Pike} J.G. McWhirter and E.R. Pike, On the numerical inversion of the
Laplace transform and similar Fredholm integral equations of the first kind, J. Phys.
A 11, 1729-1745 (1978).

\bibitem{Laplace} N.Ostrowsky, D.Sornette, P.Parker and E.R.Pike,
Exponential sampling method for light scattering polydispersity analysis,  Optica Acta
28,1059 (1981).

\bibitem{Chaline} J. Chaline, L. Nottale and P. Grou, L'arbre de la vie a-t'il une 
structure math\'ematique fractale?, Preprint 1998.

\bibitem{SmithHH} C.E. Smith, editor, Log periodic antenna design handbook, 1st ed.
Cleveland, Ohio, Smith Electronics, c1966 (1979 printing)

\bibitem{Baker} D.C. Baker and T.G. Reuss, An
investigation of the design of a log-periodic dipole array with low side-lobe levels
for broadcast applications,  IEEE Transactions on Broadcasting 36, 89-93 (1990)

\bibitem{Dykaar} D.R. Dykaar, B.I.
Greene, J.F. Federici, A.F.J. Levi et al., Log-periodic antennas for pulsed terahertz
radiation, Applied Physics Letters 59, 262-264 (1991)

\bibitem{Smithh} H.K. Smith and P.E. Mayes, Log-periodic array of
dual-fed microstrip path antennas, IEEE Transactions of Antennas and Propagation 39,
1659-1664 (1991)

\bibitem{Excell} P.S. Excell,, N.N. Jackson and K.T. Wong,
Compact electromagnetic test range using an array of log-periodic antennas, IEE
Proceeding-H Microwaves Antennas and Propagation 140, 101-106 (1993)

\bibitem{Delyser} R.R. Delyser, D.C. Chang and E.F.
Kuester, Design of a  log-periodic trip grating microstrip antenna, International
Journal of Microwave and Millimeter-Wave Computer-Aided Engineering 3, 143-150 (1993)

\bibitem{Gitin} M.M.
Gitin, F.W. Wise, G. Arjavalingam, Y. Pastol et al., Broad-band characterization of
millimeter-wave log-periodic antennas by photoconductive sampling, IEEE Transactions
of Antennas and Propagation 42, 335-339 (1994)

\bibitem{Derrida} B. Derrida, J. P. Eckmann, A. Erzan, J. Phys. A16 893 (1983)

\bibitem{Falconer} K.J. Falconer, J. Theoret. Prob. 7, 681 (1994).

\bibitem{Barnsley} M.F. Barnsley, Fractals everywhere, 2nd ed. (rev. with the assistance
of Hawley Rising III),  Boston\,: Academic Press Professional, 1993.

\bibitem{Erzan} A. Erzan and J.-P. Eckmann, $q$-analysis of fractal sets, Phys. Rev.
Lett. 78, 3245-3248, 1997.

\bibitem{Erzan2} A. Erzan and A. Gorbon, Mon-commutative geometry and irreversibility, 
Eur. Phys. J. B 1, 111-116 (1998).

\bibitem{Solis} F.J. Solis and L. Tao, Lacunarity of random fractals, 
Phys. Lett. A 228, 351-356 (1997) (cond-mat/9703051).

\bibitem{Luckbook} J.-M. Luck, Syst\`emes d\'esordonn\'es unidimensionnels (Al\'ea Saclay, 
Commissariat \`a l'Energie Atomique, 1992).

\bibitem{SOCtec} A.Sornette and D.Sornette, Self-organized criticality and earthquakes,
Europhys.Lett. 9, 197 (1989); D.Sornette, Ph.Davy and A. Sornette,
Structuration of the lithosphere in plate tectonics as a self-organized critical
phenomenon, J.Geophys.Res.95, 17353 (1990).

\bibitem{Pazmandi} F. Pazmandi, R.T. Scalettar and G.T. Zimanyi, Revisiting the 
theory of finite size scaling in disordered systems\,: $\nu$ can be less than ${2 \over d}$,
Phys. Rev. Lett. 79, 5130-5133 (1997).

\bibitem{Vanneste} D. Sornette  and C.Vanneste,
Dynamics and memory effects in rupture of thermal fuse networks, Phys.Rev.Lett. 
68, 612-615 (1992); D.Sornette, C.Vanneste and L.Knopoff,
Statistical model of earthquake foreshocks, Phys.Rev.A 45, 8351-8357 (1992);
C.Vanneste and D.Sornette, Dynamics of rupture in thermal fuse models,
 J.Phys.I France 2, 1621-1644 (1992); L. Lamaign\`ere, F. Carmona and D. Sornette,
Experimental realization of critical thermal fuse rupture, Phys. Rev. Lett. 77,
2738-2741 (1996).

\bibitem{finitezeiz} A. Johansen and D. Sornette,
Evidence of discrete scale invariance by canonical averaging,
Int. J. Mod. Phys. C 9, 433-447 (1998).

\bibitem{boro} F.M. Borodich, Renormalization schemes for earthquake prediction, Geophys. J.
Int. 131, 171-178 (1997); Parametric-homogeneity and self-similar phenomena, Nonlinear
Analysis-Theory Methods \& Applications 30, 409-417 (1997).

\end{thebibliography}
\end{document}